\documentclass[11pt]{IEEEtran}
\usepackage[latin9]{inputenc}
\usepackage{array}
\usepackage{varioref}
\usepackage{float}
\usepackage{amsmath}
\usepackage{amsthm}
\usepackage{amssymb}
\usepackage{graphicx}

\makeatletter

\providecommand{\tabularnewline}{\\}

\theoremstyle{plain}
\newtheorem{thm}{\protect\theoremname}
\theoremstyle{remark}
\newtheorem{rem}[thm]{\protect\remarkname}
\theoremstyle{plain}
\newtheorem{prop}[thm]{\protect\propositionname}

\usepackage{algorithm,algpseudocode}

\@ifundefined{showcaptionsetup}{}{%
 \PassOptionsToPackage{caption=false}{subfig}}
\usepackage{subfig}
\makeatother

\providecommand{\propositionname}{Proposition}
\providecommand{\remarkname}{Remark}
\providecommand{\theoremname}{Theorem}

\begin{document}

\title{A Solution for Large-scale Multi-object Tracking}

\author{Michael Beard, Ba Tuong Vo, Ba-Ngu Vo\thanks{M. Beard, B. T. Vo and B.-N. Vo are with the School of Electrical
Engineering, Computing and Mathematical Sciences, Curtin University,
Bentley, WA 6102, Australia.}}
\maketitle
\begin{abstract}
A large-scale multi-object tracker based on the generalised labeled
multi-Bernoulli (GLMB) filter is proposed. The algorithm is capable
of tracking a very large, unknown and time-varying number of objects
simultaneously, in the presence of a high number of false alarms,
as well as misdetections and measurement origin uncertainty due to
closely spaced objects. The algorithm is demonstrated on a simulated
large-scale tracking scenario, where the peak number objects appearing
simultaneously exceeds one million. To evaluate the performance of
the proposed tracker, we also introduce a new method of applying the
optimal sub-pattern assignment (OSPA) metric, and an efficient strategy
for its evaluation in large-scale scenarios.
\end{abstract}


\section{Introduction}

Multi-object tracking has a host of applications in diverse areas,
and many effective solutions have been developed in the literature
\cite{Blackman1999,Bar-Shalom1995,Mahler2007}. Due to the unknown
and time-varying number of objects, clutter, misdetection and data
association uncertainty, multi-object tracking is computationally
intensive, specifically, it is an NP-hard problem \cite{Blackman1999}.

The total number of objects/tracks, however, is not indicative of
the size or difficulty of the problem since it is straightforward
to track an arbitrarily large cumulative number of objects over time
in scenarios with only a few objects per frame. A more discerning
indicator of problem size is the representative number of objects
and measurements per frame, provided all other factors are on equal
footing. 

Computational requirement is the biggest barrier in large-scale multi-object
tracking problems that can arise in applications ranging across areas
such as space surveillance to cell biology. In space surveillance,
tracking orbital space objects including thousands of satellites and
millions of man-made debris is the foremost task \cite{Jones2013}.
Wide Area Surveillance for large urban environments, important for
disaster relief as well as traffic and accident management, requires
tracking hundreds of thousands of objects over time, including people
in crowded environments \cite{Reillyetal2010}. In cell biology, tracking
cells is critical to understanding how cells behave in living tissues
\cite{Chenouard2013}. Similarly, in wildlife biology, animal tracking
is needed to study wildlife behaviour in their natural habitats. Such
biological tracking applications could involve hundreds of thousands
of objects.

To the best of our knowledge, so far no multi-object trackers capable
of handling thousands of objects/measurements per frame have been
reported. The large-scale air traffic surveillance algorithm proposed
in \cite{Wang1999}, arguably one of the earliest large-scale multi-object
trackers, was demonstrated on a total of about 800 tracks (the number
of objects per frame is not provided). In \cite{Musicki2004}, the
scalability of the linear multi-target integrated probabilistic data
association (LMIPDA) filter was demonstrated on scenarios with 50
objects per frame. Multiple hypothesis tracking (MHT) based algorithms
developed for cell tracking \cite{Chenouard2013}, and for wildlife
tracking \cite{Betketal2007}, were demonstrated on thousands of objects
in total, with several hundreds of objects per frame \cite{Betketal2007}.
In \cite{Vo2014b} it was demonstrated that a one-term approximation
of the generalised labeled multi-Bernoulli (GLMB) filter can track
over a thousand objects per frame \cite{Vo2014b}.

Large-scale multi-object tracking, involving tens of thousands up
to millions of objects/measurements per frame, is extremely challenging
as the computational complexity does not scale gracefully with problem
size. There are numerous combinations of events, and this number explodes
with more objects/measurements. Thus, innovative theoretical and numerical
constructs are necessary to identify the important combinations and
process them in the most efficient manner. Given the current status
of computing technology, is it possible to solve large-scale multi-object
tracking problems in a timely manner, possibly even in real time? 

In this work, we present a large-scale multi-object filter capable
of tracking in the order of a million objects/measurements per frame,
while running off-the-shelf computing equipment. Our solution rests
on the notion of multi-object density\footnote{which is equivalent to a probability density for finite-set-valued
random variable \cite{Vo2005} } \textendash{} a trademark of the random finite set (RFS) approach
\textendash{} that encapsulates all information on the current set
of tracks in a single non-negative function. Processing the massive
number of combinations translates to recursive computation of the
multi-object filtering density \cite{Vo2013,Vo2014a,Vo2017}. Tractability
hinges on efficient functional approximation/computation of the so-called
GLMB filtering recursion, under limited processing/memory resources.
At a conceptual level, the key enablers in our proposed large-scale
tracker are: 
\begin{itemize}
\item Adaptive approximation of the GLMB filtering density, at each time,
by a product of tractable and approximately independent GLMB densities; 
\item Efficient parallel computation of these GLMB densities by exploiting
conjugacy of the GLMB family.
\end{itemize}
In essence, this strategy efficiently identifies and processes significant
combinations by exploiting structural properties of the GLMB densities
and parallelisation, to make the most of the limited computing resources.
Consequently, while the focus of this paper is on large-scale problems,
our solution also provides significant efficiency gains when applied
to smaller scale problems.

Our study would be incomplete without evaluating the tracking performance
of the proposed large-scale multi-object tracker, which is a major
barrier in itself. The standard application of the optimal sub-pattern
assignment (OSPA) metric, originally proposed in \cite{Schuhmacher2008},
does not take into account aspects such as track switching, and track
fragmentation. What we need is a measure of dissimilarity between
two sets of tracks, which; (i) is physically meaningful, (ii) satisfies
the properties of a metric for mathematical consistency, and (iii)
is computable for scenarios involving millions of tracks. Fortunately,
the OSPA metric can be adapted for this purpose, which we have called
OSPA\textsuperscript{(2)} to distinguish it from the standard use
of OSPA. To evaluate the performance of the proposed large-scale tracking
algorithm on a scenario involving an unknown and time-varying number
of objects in excess of a million per frame, we developed an efficient
and scalable procedure for computing the OSPA\textsuperscript{(2)}
metric.

\section{Background: Generalised Labeled Multi-Bernoulli Tracking Filter}

The multi-object \textit{tracking} problem is distinct from the multi-object
\textit{filtering} problem, in the sense that we are interested in
estimating the trajectories of objects over time, as opposed to the
multi-object state at each time instant. We therefore limit our attention
to methods that are capable of estimating object trajectories over
time, and not just multi-object states.

The generalised labelled multi-Bernoulli (GLMB) filter is an algorithm
that is specifically designed to fulfil this requirement. Its ability
to estimate trajectories stems from the idea of modeling the multi-object
state as a \textit{labeled} random finite set, which, from an intuitive
viewpoint, simply means augmenting the kinematic state space with
a discrete space of object labels. A given element of the label space
identifies a unique object, and the trajectory of an object can be
estimated by extracting the collection of states with the same identifying
label across different time instants. In the remainder of this section,
we briefly revisit the main points of the GLMB filter, the interested
reader is referred to \cite{Vo2013,Vo2014a,Vo2017,Reuter2014} for
a more detailed treatment.

We begin by defining the notion of a labeled random finite set. Let
$\mathbb{X}$ be a single-object state space, $\mathbb{L}$ a discrete
label space, and $\mathcal{L}:\mathbb{X}\times\mathbb{L}\rightarrow\mathbb{L}$
the projection defined by $\mathcal{L}\left(\left(x,\ell\right)\right)=\ell$
for all points $\left(x,\ell\right)\in\mathbb{X}\times\mathbb{L}$.
Consider a finite subset $\boldsymbol{X}\subseteq\mathbb{X}\times\mathbb{L}$,
and its corresponding label set $\mathcal{L}\left(\boldsymbol{X}\right)=\left\{ \mathcal{L}\left(\boldsymbol{x}\right):x\in\boldsymbol{X}\right\} $.
It follows that the labels of the points in $\boldsymbol{X}$ are
distinct if and only if $\boldsymbol{X}$ and its label set $\mathcal{L}\left(\boldsymbol{X}\right)$
have equal cardinality. This is expressed mathematically by defining
a \textit{distinct label indicator} function
\begin{align*}
\Delta\left(\boldsymbol{X}\right) & \triangleq\delta_{\left|\boldsymbol{X}\right|}\left[\left|\mathcal{L}\left(\boldsymbol{X}\right)\right|\right],
\end{align*}
which has value 1 if the labels in $\boldsymbol{X}$ are distinct
and 0 otherwise. A \textit{labeled RFS} is defined as an RFS on the
space $\mathbb{X}\times\mathbb{L}$, with the constraint that any
realisation $\boldsymbol{X}$ must satisfy $\Delta\left(\boldsymbol{X}\right)=1$.

A generalised labeled multi-Bernoulli RFS is a class of labeled RFS
that is distributed according a multi-object density with the form
\begin{align*}
\boldsymbol{\pi}\left(\boldsymbol{X}\right) & =\Delta\left(\boldsymbol{X}\right)\sum_{\xi\in\Xi}w^{\left(\xi\right)}\left(\mathcal{L}\left(\boldsymbol{X}\right)\right)\left[p^{\left(\xi\right)}\right]^{\boldsymbol{X}},
\end{align*}
where $\Xi$ is a discrete index set, each $w^{\left(\xi\right)}\left(L\right)$
is a non-negative weight such that $\sum_{\xi\in\Xi}\sum_{L\in\mathcal{F}\left(\mathbb{L}\right)}w^{\left(\xi\right)}\left(L\right)=1$,
each $p^{\left(\xi\right)}\left(\cdot,\ell\right)$ is a probability
density on $\mathbb{X}$, and $\left[h\right]^{X}\triangleq\prod_{x\in X}h\left(x\right)$
is referred to as a \textit{multi-object exponential}. As shown in
\cite{Vo2013}, the GLMB is closed under the the standard multi-object
transition kernel, and is a conjugate prior with respect to the standard
multi-object observation likelihood. These properties ensure that
the GLMB can be used to construct a recursive Bayesian multi-object
tracker.

Let us assume that at time $k-1$, the posterior multi-object density
is a GLMB for the form
\begin{align}
\boldsymbol{\pi}_{k-1}\left(\boldsymbol{X}\right) & =\Delta\left(\boldsymbol{X}\right)\sum_{I,\xi}\omega_{k-1}^{\left(I,\xi\right)}\delta_{I}\left(\mathcal{L}\left(\boldsymbol{X}\right)\right)\left[p_{k-1}^{\xi}\right]^{\boldsymbol{X}}.
\end{align}

\begin{figure*}
\rule[0.5ex]{1\textwidth}{0.5pt}

\begin{align}
\boldsymbol{\pi}_{k}\left(\boldsymbol{X}\right) & \propto\Delta\left(\boldsymbol{X}\right)\sum_{J,\xi,\theta}\left(\sum_{I}\omega_{k-1}^{\left(I,\xi\right)}\omega_{k}^{\left(I,\xi,J,\theta\right)}\left(Z_{k}\right)\right)\delta_{J}\left(\mathcal{L}\left(\boldsymbol{X}\right)\right)\left[p_{k}^{\left(\xi,\theta\right)}\left(\cdot|Z_{k}\right)\right]^{\boldsymbol{X}}\label{e:GLMB_Posterior_First}
\end{align}

where

\begin{align}
\omega_{k}^{\left(I,\xi,J,\theta\right)}\left(Z_{k}\right) & =1_{\Theta_{k}\left(J\right)}\left(\theta\right)\left[1-r_{B,k}\right]^{\mathbb{B}_{k}-J}\left[r_{B,k}\right]^{\mathbb{B}_{k}\cap J}\left[1-\bar{P}_{S,k}^{\left(\xi\right)}\right]^{I-J}\left[\bar{P}_{S,k}^{\left(\xi\right)}\right]^{I\cap J}\left[\bar{\Psi}{}_{Z_{k},k}^{\left(\xi,\theta\right)}\right]^{J}\\
\bar{\Psi}_{Z_{k},k}^{\left(\xi,\theta\right)}\left(\ell\right) & =\left\langle \Psi_{Z_{k},k}^{\left(\xi,\theta\left(\ell\right)\right)}\left(\cdot,\ell\right),p_{k|k-1}^{\left(\xi\right)}\left(\cdot,\ell\right)\right\rangle \\
\bar{P}_{S,k}^{\left(\xi\right)}\left(\ell\right) & =\left\langle P_{S,k}\left(\cdot,\ell\right),p_{k-1}^{\left(\xi\right)}\left(\cdot,\ell\right)\right\rangle \\
p_{k}^{\left(\xi,\theta\right)}\left(x,\ell|Z_{k}\right) & \propto\Psi_{Z_{k},k}^{\left(\xi,\theta\left(\ell\right)\right)}\left(x,\ell\right)p_{k|k-1}^{\left(\xi\right)}\left(x,\ell\right)\\
p_{k|k-1}^{\left(\xi\right)}\left(x,\ell\right) & =1_{\mathbb{B}_{k}}\left(\ell\right)p_{B,k}\left(x,\ell\right)+1_{\mathbb{L}_{k-1}}\left(\ell\right)\frac{\left\langle P_{S,k}\left(\cdot,\ell\right)f_{k|k-1}\left(x|\cdot,\ell\right),p_{k-1}^{\left(\xi\right)}\left(\cdot,\ell\right)\right\rangle }{\bar{P}_{S,k}^{\left(\xi\right)}\left(\ell\right)}\\
\Psi_{Z_{k},k}^{\left(\xi,\theta\left(\ell\right)\right)}\left(x,\ell\right) & =\delta\left(\theta\left(\ell\right)\right)\left(1-P_{D,k}\left(x,\ell\right)\right)+\left(1-\delta\left(\theta\left(\ell\right)\right)\right)\frac{P_{D,k}\left(x,\ell\right)g\left(z_{\theta\left(\ell\right)}|x,\ell\right)}{\kappa\left(z_{\theta\left(\ell\right)}\right)}\label{e:GLMB_Posterior_Last}
\end{align}

\rule[0.5ex]{1\textwidth}{0.5pt}
\end{figure*}

\begin{table}[H]
\vspace{0.75cm}
\begin{centering}
\begin{tabular}{|l|>{\raggedright}p{6cm}|}
\hline 
\textbf{Symbol} & \textbf{Definition}\tabularnewline
\hline 
\hline 
$r_{B,k}\left(\ell\right)$ & Probability that an object with label $\ell$ is born at time $k$\tabularnewline
\hline 
$p_{B,k}$$\left(\cdot,\ell\right)$ & Prior probability density for a new object born at time $k$ with
label $\ell$\tabularnewline
\hline 
$\mathbb{B}_{k}$ & Label space for new objects born at time $k$\tabularnewline
\hline 
$\mathbb{L}_{k-1}$ & Space of all possible object labels up to time $k-1$ \tabularnewline
\hline 
$Z_{k}$ & Set of measurements received at time $k$\tabularnewline
\hline 
$\Theta_{k}$ & Space of measurement-label mappings at time $k$\tabularnewline
\hline 
$\theta\left(\ell\right)$ & Measurement index assigned to label $\ell$ under mapping $\theta\in\Theta_{k}$,
where $\theta\left(\ell\right)=0$ implies that no measurement is
assigned\tabularnewline
\hline 
$P_{S,k}\left(x,\ell\right)$ & Probability that an object with state $x$ and label $\ell$ at time
$k-1$ survives to time $k$\tabularnewline
\hline 
$P_{D,k}$$\left(x,\ell\right)$ & Probability that an object with state $x$ and label $\ell$ is detected
at time $k$\tabularnewline
\hline 
$f_{k|k-1}\left(\cdot|x,\ell\right)$ & Prior probability density for a surviving object at time $k$, which
had state $x$ and label $\ell$ at time $k-1$\tabularnewline
\hline 
$g\left(\cdot|x,\ell\right)$ & Probability density for an observation generated by an object with
state $x$ and label $\ell$\tabularnewline
\hline 
\end{tabular}
\par\end{centering}
\caption{Definitions of symbols used in GLMB filter equations \eqref{e:GLMB_Posterior_First}-\eqref{e:GLMB_Posterior_Last}}
\label{t:GLMB_Symbols}
\end{table}

Then at time $k$, the posterior multi-object density is given by
the GLMB as defined by equations \eqref{e:GLMB_Posterior_First}-\eqref{e:GLMB_Posterior_Last},
where we have used the notation defined in Table \eqref{t:GLMB_Symbols}
to denote the inputs and model parameters.

\section{Scalable GLMB Filtering: Theoretical Foundations}

Due to practical limitations on computational resources, the original
implementation of the GLMB filter proposed in \cite{Vo2013,Vo2014a}
cannot accommodate large numbers of objects simultaneously. The main
computational bottleneck occurs in the measurement update procedure,
which involves processing each component of the predicted GLMB density
using Murty's algorithm to rank the most significant posterior components
in descending order of weight. Accordingly, if the input data consists
of $M$ measurements, and the $i$-th component of the predicted GLMB
density has $N$ unique target labels and is allocated $K$ posterior
components, then the complexity of processing the $i$-th component
will be $\mathcal{O}\left(K\left(N+M\right)^{3}\right)$. Clearly
the number of measurements and objects that this algorithm can handle
in practice will depend on the available resources. However, in general,
the cubic complexity will render this algorithm computationally infeasible
for large values of $N$ and $M$.

An alternative implementation of the GLMB filter based on Gibbs sampling
was proposed in \cite{Vo2017}. This algorithm replaces the expensive
procedure of deterministically generating the highest weighted components,
in favour of a cheaper stochastic sampling approach. Consequently,
this method reduces the computational complexity of processing each
component to $\mathcal{O}\left(KN^{2}M\right)$. Note that the quadratic
complexity in the number of objects will still render this algorithm
infeasible for tracking large numbers of objects simultaneously.

In a given application, the available computational resources, combined
with the requirements on processing speed and tracking accuracy, will
impose a practical upper limit on the number of objects that can be
feasibly tracked simultaneously. In order to track more objects using
this algorithm, the overall tracking problem needs to be decomposed
into smaller sub-problems, with an upper bound on the number of objects
in any given sub-problem. In labelled multi-object tracking, such
a decomposition can be described in terms of a partition of the label
space, where each group within the partition represents a sub-problem
that is assumed to be statistically independent of all others. For
a given a partition of the label space, it is possible to approximate
the overall GLMB density as a product of smaller GLMB densities, each
of which consists only of labels from a single group within the partition.
Here we present a method for computing these GLMB densities, which
is optimal in the sense that the Kullback-Leibler divergence (KLD)
from the original GLMB density is minimised. In the remainder of this
section we show that, for a given partition of the label space, a
GLMB density can be feasibly and optimally decomposed into a product
of smaller GLMB densities.

\subsection{GLMB Optimal Product Approximation}

Given a partition $G=\left\{ \mathbb{L}^{\left(1\right)},\dots,\mathbb{L}^{\left(N\right)}\right\} $
of the label space $\mathbb{L}$, i.e. $\mathbb{L}=\biguplus\limits _{n=1}^{N}\mathbb{L}^{\left(n\right)}$,
then
\[
\mathcal{F}\left(\mathbb{X}\times\mathbb{L}\right)=\biguplus\limits _{n=1}^{N}\mathcal{F}\left(\mathbb{X}\times\mathbb{L}^{\left(n\right)}\right).
\]
Let $\boldsymbol{X}^{(n)}=\boldsymbol{X}\cap\mathcal{F}\left(\mathbb{X}\times\mathbb{L}^{\left(n\right)}\right)$,
then
\[
\boldsymbol{X}=\biguplus\limits _{n=1}^{N}\boldsymbol{X}^{(n)},
\]
and the marginal $\boldsymbol{\pi}^{\left(n\right)}$ of $\boldsymbol{\pi}$
on $\mathcal{F}\left(\mathbb{X}\times\mathbb{L}^{\left(n\right)}\right)$
is given by
\[
\boldsymbol{\pi}^{\left(n\right)}\!\left(\!\boldsymbol{X}^{\left(n\right)}\!\right)=\int\!\boldsymbol{\pi}\!\left(\biguplus\limits _{j=1}^{N}\boldsymbol{X}^{\left(j\right)}\!\right)\!\delta\!\left(\!\boldsymbol{X}^{\left(1:n-1\right)},\boldsymbol{X}^{\left(n+1:N\right)}\!\right)
\]
We are interested in approximating $\boldsymbol{\pi}$ as a product
of $N$ independent densities as follows
\[
\boldsymbol{\pi}\left(\biguplus\limits _{n=1}^{N}\boldsymbol{X}^{\left(n\right)}\right)\cong\prod_{n=1}^{N}\tilde{\boldsymbol{\pi}}^{\left(n\right)}\left(\boldsymbol{X}^{\left(n\right)}\right),
\]
where $\tilde{\boldsymbol{\pi}}^{\left(n\right)}\left(\cdot\right)$
is a density on the space $\mathcal{F}\left(\mathbb{X}\times\mathbb{L}^{\left(n\right)}\right)$.
First let us consider a partition consisting of two groups i.e. $G=\left\{ \mathbb{L}^{\left(1\right)},\mathbb{L}^{\left(2\right)}\right\} $
. The Kullback-Leibler divergence between $\boldsymbol{\pi}$ and
its product approximation is given by \eqref{e:KLD_Product}\vpageref{e:KLD_Product}.
Note that the minimum possible KLD is obtained when $D\left(\boldsymbol{\pi}^{\left(1\right)};\tilde{\boldsymbol{\pi}}^{\left(1\right)}\right)=0$
and $D\left(\boldsymbol{\pi}^{\left(2\right)};\tilde{\boldsymbol{\pi}}^{\left(2\right)}\right)=0$,
which, by the properties of the KLD, occurs only when $\tilde{\boldsymbol{\pi}}^{\left(1\right)}=\boldsymbol{\pi}^{\left(1\right)}$
and $\tilde{\boldsymbol{\pi}}^{\left(2\right)}=\boldsymbol{\pi}^{\left(2\right)}$
almost everywhere. Thus, choosing the elements of the product approximation
to be the marginal densities yields the minimum possible KLD. The
same argument follows for a general partition consisting of $N$ groups,
i.e. $G=\left\{ \mathbb{L}^{\left(1\right)},\dots,\mathbb{L}^{\left(N\right)}\right\} $.

\begin{figure*}
\rule[0.5ex]{1\textwidth}{0.5pt}

\begin{align}
 & D\left(\boldsymbol{\pi};\tilde{\boldsymbol{\pi}}^{\left(1\right)}\tilde{\boldsymbol{\pi}}^{\left(2\right)}\right)\nonumber \\
 & \qquad=\int\int\log\left(\frac{\boldsymbol{\pi}\left(\boldsymbol{X}^{\left(1\right)}\uplus\boldsymbol{X}^{\left(2\right)}\right)}{\tilde{\boldsymbol{\pi}}^{\left(1\right)}\left(\boldsymbol{X}^{\left(1\right)}\right)\tilde{\boldsymbol{\pi}}^{\left(2\right)}\left(\boldsymbol{X}^{\left(2\right)}\right)}\right)\boldsymbol{\pi}\left(\boldsymbol{X}^{\left(1\right)}\uplus\boldsymbol{X}^{\left(2\right)}\right)\delta\boldsymbol{X}^{\left(1\right)}\delta\boldsymbol{X}^{\left(2\right)}\nonumber \\
 & \qquad=\int\int\log\left(\frac{\boldsymbol{\pi}\left(\boldsymbol{X}^{\left(1\right)}\uplus\boldsymbol{X}^{\left(2\right)}\right)}{\boldsymbol{\pi}^{\left(1\right)}\left(\boldsymbol{X}^{\left(1\right)}\right)\boldsymbol{\pi}^{\left(2\right)}\left(\boldsymbol{X}^{\left(2\right)}\right)}\frac{\boldsymbol{\pi}^{\left(1\right)}\left(\boldsymbol{X}^{\left(1\right)}\right)\boldsymbol{\pi}^{\left(2\right)}\left(\boldsymbol{X}^{\left(2\right)}\right)}{\tilde{\boldsymbol{\pi}}^{\left(1\right)}\left(\boldsymbol{X}^{\left(1\right)}\right)\tilde{\boldsymbol{\pi}}^{\left(2\right)}\left(\boldsymbol{X}^{\left(2\right)}\right)}\right)\boldsymbol{\pi}\left(\boldsymbol{X}^{\left(1\right)}\uplus\boldsymbol{X}^{\left(2\right)}\right)\delta\boldsymbol{X}^{\left(1\right)}\delta\boldsymbol{X}^{\left(2\right)}\nonumber \\
 & \qquad=\int\int\log\left(\frac{\boldsymbol{\pi}\left(\boldsymbol{X}^{\left(1\right)}\uplus\boldsymbol{X}^{\left(2\right)}\right)}{\boldsymbol{\pi}^{\left(1\right)}\left(\boldsymbol{X}^{\left(1\right)}\right)\boldsymbol{\pi}^{\left(2\right)}\left(\boldsymbol{X}^{\left(2\right)}\right)}\right)\boldsymbol{\pi}\left(\boldsymbol{X}^{\left(1\right)}\uplus\boldsymbol{X}^{\left(2\right)}\right)\delta\boldsymbol{X}^{\left(1\right)}\delta\boldsymbol{X}^{\left(2\right)}\nonumber \\
 & \qquad\qquad+\int\int\log\left(\frac{\boldsymbol{\pi}^{\left(1\right)}\left(\boldsymbol{X}^{\left(1\right)}\right)\boldsymbol{\pi}^{\left(2\right)}\left(\boldsymbol{X}^{\left(2\right)}\right)}{\tilde{\boldsymbol{\pi}}^{\left(1\right)}\left(\boldsymbol{X}^{\left(1\right)}\right)\tilde{\boldsymbol{\pi}}^{\left(2\right)}\left(\boldsymbol{X}^{\left(2\right)}\right)}\right)\boldsymbol{\pi}\left(\boldsymbol{X}^{\left(1\right)}\uplus\boldsymbol{X}^{\left(2\right)}\right)\delta\boldsymbol{X}^{\left(1\right)}\delta\boldsymbol{X}^{\left(2\right)}\nonumber \\
 & \qquad=D\left(\boldsymbol{\pi};\boldsymbol{\pi}^{\left(1\right)}\boldsymbol{\pi}^{\left(2\right)}\right)+\int\int\log\left(\frac{\boldsymbol{\pi}^{\left(1\right)}\left(\boldsymbol{X}^{\left(1\right)}\right)\boldsymbol{\pi}^{\left(2\right)}\left(\boldsymbol{X}^{\left(2\right)}\right)}{\tilde{\boldsymbol{\pi}}^{\left(1\right)}\left(\boldsymbol{X}^{\left(1\right)}\right)\tilde{\boldsymbol{\pi}}^{\left(2\right)}\left(\boldsymbol{X}^{\left(2\right)}\right)}\right)\boldsymbol{\pi}\left(\boldsymbol{X}^{\left(1\right)}\uplus\boldsymbol{X}^{\left(2\right)}\right)\delta\boldsymbol{X}^{\left(1\right)}\delta\boldsymbol{X}^{\left(2\right)}\nonumber \\
 & \qquad=D\left(\boldsymbol{\pi};\boldsymbol{\pi}^{\left(1\right)}\boldsymbol{\pi}^{\left(2\right)}\right)+\int\int\log\left(\frac{\boldsymbol{\pi}^{\left(1\right)}\left(\boldsymbol{X}^{\left(1\right)}\right)}{\tilde{\boldsymbol{\pi}}^{\left(1\right)}\left(\boldsymbol{X}^{\left(1\right)}\right)}\right)\boldsymbol{\pi}\left(\boldsymbol{X}^{\left(1\right)}\uplus\boldsymbol{X}^{\left(2\right)}\right)\delta\boldsymbol{X}^{\left(1\right)}\delta\boldsymbol{X}^{\left(2\right)}\nonumber \\
 & \qquad\qquad+\int\int\log\left(\frac{\boldsymbol{\pi}^{\left(2\right)}\left(\boldsymbol{X}^{\left(2\right)}\right)}{\tilde{\boldsymbol{\pi}}^{\left(2\right)}\left(\boldsymbol{X}^{\left(2\right)}\right)}\right)\boldsymbol{\pi}\left(\boldsymbol{X}^{\left(1\right)}\uplus\boldsymbol{X}^{\left(2\right)}\right)\delta\boldsymbol{X}^{\left(1\right)}\delta\boldsymbol{X}^{\left(2\right)}\nonumber \\
 & \qquad=D\left(\boldsymbol{\pi};\boldsymbol{\pi}^{\left(1\right)}\boldsymbol{\pi}^{\left(2\right)}\right)+\int\log\left(\frac{\boldsymbol{\pi}^{\left(1\right)}\left(\boldsymbol{X}^{\left(1\right)}\right)}{\tilde{\boldsymbol{\pi}}^{\left(1\right)}\left(\boldsymbol{X}^{\left(1\right)}\right)}\right)\boldsymbol{\pi}^{\left(1\right)}\left(\boldsymbol{X}^{\left(1\right)}\right)\delta\boldsymbol{X}^{\left(1\right)}\nonumber \\
 & \qquad\qquad+\int\log\left(\frac{\boldsymbol{\pi}^{\left(2\right)}\left(\boldsymbol{X}^{\left(2\right)}\right)}{\tilde{\boldsymbol{\pi}}^{\left(2\right)}\left(\boldsymbol{X}^{\left(2\right)}\right)}\right)\boldsymbol{\pi}^{\left(2\right)}\left(\boldsymbol{X}^{\left(2\right)}\right)\delta\boldsymbol{X}^{\left(2\right)}\nonumber \\
 & \qquad=D\left(\boldsymbol{\pi};\boldsymbol{\pi}^{\left(1\right)}\boldsymbol{\pi}^{\left(2\right)}\right)+D\left(\boldsymbol{\pi}^{\left(1\right)};\tilde{\boldsymbol{\pi}}^{\left(1\right)}\right)+D\left(\boldsymbol{\pi}^{\left(2\right)};\tilde{\boldsymbol{\pi}}^{\left(2\right)}\right)\label{e:KLD_Product}
\end{align}

\rule[0.5ex]{1\textwidth}{0.5pt}
\end{figure*}

\subsection{Marginalising GLMB Densities}

Consider a GLMB density of the form
\begin{equation}
\boldsymbol{\pi}\left(\boldsymbol{X}\right)=\Delta\left(\boldsymbol{X}\right)\sum_{\xi\in\Xi}w^{\left(\xi\right)}\left(\mathcal{L}\left(\boldsymbol{X}\right)\right)\left[p^{\left(\xi\right)}\right]^{\boldsymbol{X}}.\label{e:GLMB}
\end{equation}
For a partition $G=\left\{ \mathbb{L}^{\left(1\right)},\mathbb{L}^{\left(2\right)}\right\} $
consisting of two groups, the marginalised GLMB density with respect
to the first group is given by \eqref{e:GLMB_Marginal}-\eqref{e:GLMB_Marginal_Weight}\vpageref{e:GLMB_Marginal}.
For a general partition $G=\left\{ \mathbb{L}^{\left(1\right)},\dots,\mathbb{L}^{\left(N\right)}\right\} $,
using the same argument, the marginalised density corresponding to
the $n$-th group is given by
\begin{equation}
\boldsymbol{\pi}^{\left(n\right)}\!\left(\!\boldsymbol{X}^{\left(n\right)}\!\right)=\Delta\!\left(\!\boldsymbol{X}^{\left(n\right)}\!\right)\sum_{\xi\in\Xi}w^{\left(\xi,n\right)}\!\left(\!\mathcal{L}\!\left(\!\boldsymbol{X}^{\left(n\right)}\!\right)\right)\!\left[p^{\left(\xi\right)}\right]^{\boldsymbol{X}^{\left(n\right)}}
\end{equation}
where
\[
w^{\left(\xi,n\right)}\left(I^{\left(n\right)}\right)=\sum_{I\in\mathcal{F}\left(\mathbb{L}\right)-\mathcal{F}\left(\mathbb{L}^{\left(n\right)}\right)}w^{\left(\xi\right)}\left(I^{\left(n\right)}\uplus I\right).
\]

\begin{figure*}[t]
\rule[0.5ex]{1\textwidth}{0.5pt}

\begin{align}
\boldsymbol{\pi}^{\left(1\right)}\left(\boldsymbol{X}^{\left(1\right)}\right) & =\int\Delta\left(\boldsymbol{X}^{\left(1\right)}\uplus\boldsymbol{X}^{\left(2\right)}\right)\sum_{\xi\in\Xi}w^{\left(\xi\right)}\left(\mathcal{L}\left(\boldsymbol{X}^{\left(1\right)}\uplus\boldsymbol{X}^{\left(2\right)}\right)\right)\left[p^{\left(\xi\right)}\right]^{\boldsymbol{X}^{\left(1\right)}\uplus\boldsymbol{X}^{\left(2\right)}}\delta\boldsymbol{X}^{\left(2\right)}\nonumber \\
 & =\int\Delta\left(\boldsymbol{X}^{\left(1\right)}\right)\Delta\left(\boldsymbol{X}^{\left(2\right)}\right)\sum_{\xi\in\Xi}w^{\left(\xi\right)}\left(\mathcal{L}\left(\boldsymbol{X}^{\left(1\right)}\right)\uplus\mathcal{L}\left(\boldsymbol{X}^{\left(2\right)}\right)\right)\left[p^{\left(\xi\right)}\right]^{\boldsymbol{X}^{\left(1\right)}}\left[p^{\left(\xi\right)}\right]^{\boldsymbol{X}^{\left(2\right)}}\delta\boldsymbol{X}^{\left(2\right)}\nonumber \\
 & =\Delta\left(\boldsymbol{X}^{\left(1\right)}\right)\sum_{\xi\in\Xi}w^{\left(\xi,1\right)}\left(\mathcal{L}\left(\boldsymbol{X}^{\left(1\right)}\right)\right)\left[p^{\left(\xi\right)}\right]^{\boldsymbol{X}^{\left(1\right)}}\label{e:GLMB_Marginal}
\end{align}
where
\begin{align}
w^{\left(\xi,1\right)}\left(I^{\left(1\right)}\right) & =\int\Delta\left(\boldsymbol{X}^{\left(2\right)}\right)w^{\left(\xi\right)}\left(I^{(1)}\uplus\mathcal{L}\left(\boldsymbol{X}^{\left(2\right)}\right)\right)\left[p^{(\xi)}\right]^{\boldsymbol{X}^{\left(2\right)}}\delta\boldsymbol{X}^{\left(2\right)}\nonumber \\
 & =\sum_{i=0}^{\infty}\frac{1}{i!}\sum_{\left(\ell_{1}^{\left(2\right)},\dots,\ell_{i}^{\left(2\right)}\right)\in\left(\mathbb{L}^{\left(2\right)}\right)^{i}}\int_{\mathbb{X}^{i}}\delta_{i}\left(\left|\left\{ \ell_{1}^{\left(2\right)},\dots,\ell_{i}^{\left(2\right)}\right\} \right|\right)\times\nonumber \\
 & \qquad\qquad w^{(\xi)}\left(I^{\left(1\right)}\uplus\left\{ \ell_{1}^{\left(2\right)},\dots,\ell_{i}^{\left(2\right)}\right\} \right)\left[p^{(\xi)}\right]^{\left\{ \left(x_{1}^{\left(2\right)},\ell_{1}^{\left(2\right)}\right),\dots,\left(x_{i}^{\left(2\right)},\ell_{i}^{\left(2\right)}\right)\right\} }d\left(x_{1}^{\left(2\right)},\dots,x_{i}^{\left(2\right)}\right)\nonumber \\
 & =\sum_{i=0}^{\infty}\frac{1}{i!}\sum_{\left(\ell_{1}^{(2)},\dots,\ell_{i}^{(2)}\right)\in\left(\mathbb{L}^{\left(2\right)}\right)^{i}}\delta_{i}\left(\left|\left\{ \ell_{1}^{\left(2\right)},\dots,\ell_{i}^{\left(2\right)}\right\} \right|\right)\times\nonumber \\
 & \qquad\qquad w^{\left(\xi\right)}\left(I^{\left(1\right)}\uplus\left\{ \ell_{1}^{\left(2\right)},\dots,\ell_{i}^{\left(2\right)}\right\} \right)\int_{\mathbb{X}^{i}}\left[p^{\left(\xi\right)}\right]^{\left\{ \left(x_{1}^{\left(2\right)},\ell_{1}^{\left(2\right)}\right),\dots,\left(x_{i}^{\left(2\right)},\ell_{i}^{\left(2\right)}\right)\right\} }d\left(x_{1}^{\left(2\right)},\dots,x_{i}^{\left(2\right)}\right)\nonumber \\
 & =\sum_{i=0}^{\infty}\frac{1}{i!}\sum_{\left(\ell_{1}^{(2)},\dots,\ell_{i}^{(2)}\right)\in\left(\mathbb{L}^{\left(2\right)}\right)^{i}}\delta_{i}\left(\left|\left\{ \ell_{1}^{\left(2\right)},\dots,\ell_{i}^{\left(2\right)}\right\} \right|\right)w^{\left(\xi\right)}\left(I^{\left(1\right)}\uplus\left\{ \ell_{1}^{\left(2\right)},\dots,\ell_{i}^{\left(2\right)}\right\} \right)\nonumber \\
 & =\sum_{I^{\left(2\right)}\subseteq\mathbb{L}^{\left(2\right)}}w^{\left(\xi\right)}\left(I^{\left(1\right)}\uplus I^{\left(2\right)}\right)\label{e:GLMB_Marginal_Weight}
\end{align}

\rule[0.5ex]{1\textwidth}{0.5pt}
\end{figure*}

\section{Scalable GLMB Filtering: Implementation}

In this section we discuss the implementation of an efficient and
scalable GLMB filter, using the decomposition method presented in
the previous section. It is well known that the computational complexity
of the multi-object tracking problem rises sharply as the number of
objects and/or measurements increases. This leads to the problem becoming
infeasible when the number of objects/measurements present at a given
time exceeds a certain threshold, which depends on the available computing
resources. Adding more memory and processing power to the system may
allow it to track more objects, however, this does not solve the fundamental
computational problem, and will only serve to delay its onset. A logical
approach to tracking a very large number of objects simultaneously
is to employ a divide-and-conquer strategy, by decomposing the large
tracking problem into many smaller sub-problems, each requiring significantly
fewer resources to solve. However, this introduces a new challenge,
namely, to find an appropriate decomposition of the overall problem
that can be considered ``optimal'' in some sense. The meaning of
``optimal'' in this context is difficult to define, since there
are multiple competing objectives that can influence the suitability
of a solution.

Arguably, the main objective in decomposing a large tracking problem
is to divide the set of tracks into smaller groups (i.e. partitioning
the discrete set of track labels), such that the number of tracks
in any one group does not exceed some upper bound. This is necessary
to ensure that the processing of any single group does not consume
an excessive or disproportionate share of the available computing
resources. The number of tracks in a group determines the amount of
memory required to store the parameters of the corresponding GLMB
density, and the processing time needed to perform the GLMB update.
Therefore, setting an upper bound on the group size requires careful
consideration of several factors, such as the amount of available
system memory, the CPU speed, and the real-time constraints on processing
each frame of incoming data.

Another objective is to minimise the errors incurred from carrying
out the decomposition. In this case, we wish to approximate a GLMB
density that encompasses all tracks in a scenario, by a product of
marginal GLMB densities, each of which encompasses only a subset of
the tracks. Ideally, we would like this product of marginal GLMBs
to match the original (full) GLMB density as closely as possible.
This is equivalent to minimising the Kullback-Leibler divergence between
the two representations.

Finding the ideal decomposition thus involves solving a multi-objective
optimisation problem over the space of partitions of object labels.
In theory, this problem could be cast as a constrained optimisation,
where the objective is to find a partition minimising the KLD between
the full GLMB and its product approximation, subject to a constraint
on the maximum size of any one group. However, this optimisation problem
is much harder than the original tracking problem, since the space
of label partitions grows super-exponentially as the number of tracks
increases\footnote{The number of partitions of $n$ objects is called the $n$-th Bell
number, which can be computed using the recurrence relation $B_{n}=\sum_{k=0}^{n-1}\binom{n-1}{k}B_{k}$}. Therefore, to establish a feasible decomposition algorithm, we re-cast
the problem into a form that can be solved with existing numerical
techniques.

Instead of directly attempting to perform a constrained optimisation
of the KLD over the space of partitions (which is clearly intractable),
we aim to find a partition of the labels such that any pair of objects
that do not appear in the same group are approximately statistically
independent. In the standard multi-object tracking model, with point
detections and independent object motion, this statistical dependence
arises solely via the uncertainty in the unknown association between
measurements and objects. That is, if multiple objects could have
produced a particular detection, then there is a statistical dependence
between those tracks in the posterior multi-object density. It follows
that tracks that are well separated in the measurement space will
have low posterior statistical dependence, because the probability
that these tracks give rise to closely spaced measurements is extremely
low. In this context, ``well separated'' means that the distance
between tracks in the measurement space is large compared to the measurement
noise and the uncertainty in the object's predicted location. This
is the key property that we exploit in order to find a partition of
the track labels that minimises the amount of potential ``measurement
sharing'' between objects in different groups.

To achieve this, we first project the marginal probability density
corresponding to each object label onto the measurement space, which
is used to construct a ``bounding region'' in the measurement space
for each object. We then attempt to find a partition that minimises
the amount of overlap between the bounding regions of objects in different
groups. This reduces the chances of measurement sharing, and consequently
reduces the statistical dependence between groups of objects in the
posterior GLMB density. This is based on the premise that if there
are statistically independent subsets of track labels present, then
then the GLMB density over the space of all labels can be expressed
as a product of marginal GLMBs on these subsets, without any loss
of information.

It should be noted that in the worst case, it may not be possible
to find independent subsets of track labels. In such cases, the techniques
proposed here cannot effectively reduce the computational complexity.
However, in many practical tracking scenarios, such independent subsets
do exist, and it is this aspect of the problem structure that we exploit
in order to make the most efficient use of the available computing
resources.

A naive approach to minimising the overlap between target groups would
require computing the overlap between the bounding regions for all
pairs of objects. Doing so would lead to a computational complexity
of $\mathcal{O}\left(N^{2}\right)$, making it computationally infeasible
for large-scale tracking problems involving thousands or millions
of objects. Fortunately, spatial searching algorithms can be readily
applied to this problem in order to reduce the complexity to $O(N\log N)$,
thus making it suitable for large-scale multi-object tracking. In
particular, data structures based on R-trees \cite{Guttman1984,Manolopoulos2010}
can be used to efficiently search for overlapping hyper-rectangles
in the measurement space. The sizes of these hyper-rectangles are
chosen probabilistically, such that the measurement of an object has
at least probability $P_{G}$ of falling inside the rectangle. When
overlaps between hyper-rectangles are detected, the object labels
corresponding to those bounding regions are placed in the same group
within the partition. If it is not possible to satisfy the upper bound
on the group size for a given value of $P_{G}$, the procedure is
repeated with a smaller value of $P_{G}$. This results in a more
aggressive, albeit more approximate, partitioning of the track labels.

Once the partition of the labels has been constructed, we proceed
to compute a factorised representation the posterior GLMB density.
Note that when processing each new frame of measurements, we must
begin with a factorised prior GLMB density, in which each term corresponds
to a group of labels in the partition from the previous measurement
update. The factorised posterior GLMB is computed from this by the
following three-step procedure.
\begin{enumerate}
\item Carry out a further factorisation of each term in the existing prior
GLMB density, such that no single factor contains labels from multiple
groups of the new partition.
\item Multiply subsets of these factorised densities together, so as to
reconstruct the marginal prior GLMB for each group of the new partition.
\item Compute the posterior marginal GLMB density for each group of the
new partition in parallel, using the joint prediction/update method
based on Gibbs sampling as proposed in \cite{Vo2017}.
\end{enumerate}
Note that in step 3 above, the posterior densities can be computed
in parallel because the groups are all approximately independent.
The parallelisation of this final step is the key element that leads
to a potentially very significant speedup, particularly when the algorithm
is deployed on multi-core architectures with many concurrent threads
of execution.

\section{Performance Evaluation}

\subsection{Application of the OSPA Metric to Evaluate Tracking Performance}

In \cite{Schuhmacher2008}, the optimal sub-pattern assignment (OSPA)
distance was proposed as a mathematically consistent metric for measuring
the distance between two sets of points. This has found widespread
application in the literature to evaluate the performance of multi-target
filters, where it is usually presented as a plot of the OSPA distance
between the estimated and true multi-target states at each time instant.
This technique can also provide an indication of the performance of
multi-target trackers, however, it does not fully account for errors
between the estimated and true sets of tracks. When evaluating tracking
performance, the main drawback of using the OSPA distance between
multi-target states is that phenomena such as track switching and
fragmentation are not penalised in a consistent fashion.

In \cite{Beard2017}, it was shown that the OSPA metric can be applied
in a relatively straightforward manner to facilitate a more rigorous
evaluation of multi-target tracking performance. The resulting metric
is referred to as OSPA\textsuperscript{(2)}, which reflects the fact
that it is the same as the original OSPA metric, except that the base
distance is itself an OSPA-based distance. In the following, we revisit
the main points of the OSPA\textsuperscript{(2)} metric, and demonstrate
its application to the evaluation of large-scale multi-target tracking
performance. We begin by defining the following notation:
\begin{itemize}
\item $\mathbb{T}=\left\{ 1,2,\dots,K\right\} $ is a finite space of time
indices, which includes all time indices from the beginning to the
end of the scenario.
\item $\mathbb{X}$ is the single-object state space, and $\mathcal{F}\left(\mathbb{X}\right)$
is the space of finite subsets of $\mathbb{X}$.
\item $\mathbb{U}$ is the space of all functions mapping time indices in
$\mathbb{T}$ to state vectors in $\mathbb{X}$, i.e. $\mathbb{U}=\left\{ f:\mathbb{T}\mapsto\mathbb{X}\right\} $.
We refer to each element of $\mathbb{U}$ as a \textit{track}.
\item For any $f\in\mathbb{U}$, its domain, denoted by $\mathcal{D}_{f}\subseteq\mathbb{T}$,
is the set of time instants at which the object exists.
\end{itemize}
Also, recall that for a function $d\left(x,y\right)$ to be called
a ``metric'', it must satisfy the following four properties:
\begin{enumerate}
\item[P1.] $d\left(x,y\right)\geq0$ (non-negativity),
\item[P2.] $d\left(x,y\right)=0\iff x=y$ (identity),
\item[P3.] $d\left(x,y\right)=d\left(y,x\right)$ (symmetry),
\item[P4.] $d\left(x,y\right)\leq d\left(x,z\right)+d\left(z,y\right)$ (triangle
inequality).
\end{enumerate}
As defined in \cite{Schuhmacher2008}, let $d_{p}^{\left(c\right)}\left(\phi,\psi\right)$
be the OSPA distance between $\phi,\psi\in\mathcal{F}\left(\mathbb{X}\right)$
with order $p$ and cutoff $c$. That is, for $\phi=\left\{ \phi^{\left(1\right)},\phi^{\left(2\right)},\dots,\phi^{\left(m\right)}\right\} $
and $\psi=\left\{ \psi^{\left(1\right)},\psi^{\left(2\right)},\dots,\psi^{\left(n\right)}\right\} $,
with $m\leq n$
\begin{align}
 & d_{p}^{\left(c\right)}\left(\phi,\psi\right)\label{e:OSPA_Distance}\\
 & =\!\left(\!\frac{1}{n}\!\left(\!\min_{\pi\in\Pi_{n}}\sum_{i=1}^{m}\bar{d}^{\left(c\right)}\!\left(\!\phi^{\left(i\right)},\psi^{\left(\pi\left(i\right)\right)}\!\right)^{p}\!+c^{p}\left(n-m\right)\!\right)\!\right)^{1/p}\nonumber 
\end{align}
where $\bar{d}^{\left(c\right)}\left(\phi^{\left(i\right)},\psi^{\left(i\right)}\right)=\min\left(c,d\left(\phi^{\left(i\right)},\psi^{\left(i\right)}\right)\right)$,
in which $d\left(\cdot,\cdot\right)$ is a metric on the single-object
state space $\mathbb{X}$. If $m>n$, then $d_{p}^{\left(c\right)}\left(\phi,\psi\right)\triangleq d_{p}^{\left(c\right)}\left(\psi,\phi\right)$. 
\begin{rem}
Note that in \eqref{e:OSPA_Distance}, the factor of $1/n$, which
normalises the distance by the number of objects, is crucial for the
OSPA to have the intuitive interpretation as a per-object error.
\end{rem}

\subsubsection{Base Distance Between Tracks \label{s:Track_Base_Distance}}

We shall now make use of the OSPA distance \eqref{e:OSPA_Distance}
to define a metric on the space of tracks $\mathbb{U}$, which shall
in turn serve as the base distance for the OSPA\textsuperscript{(2)}
metric on the space of finite sets of tracks $\mathcal{F}\left(\mathbb{U}\right)$.
Let us define the distance $\tilde{d}^{\left(c\right)}\left(x,y\right)$
between two tracks $x,y\in\mathbb{U}$ as the mean OSPA distance between
the set of states defined by $x$ and $y$, over all times $t\in\mathcal{D}_{x}\cup\mathcal{D}_{y}$,
i.e.
\begin{align}
\tilde{d}^{\left(c\right)}\!\left(x,y\right)= & \begin{cases}
\sum\limits _{t\in\mathcal{D}_{x}\cup\mathcal{D}_{y}}\!\frac{d^{\left(c\right)}\left(\left\{ x\left(t\right)\right\} ,\left\{ y\left(t\right)\right\} \right)}{\left|\mathcal{D}_{x}\cup\mathcal{D}_{y}\right|}, & \!\!\mathcal{D}_{x}\cup\mathcal{D}_{y}\neq\emptyset\\
0, & \!\!\mathcal{D}_{x}\cup\mathcal{D}_{y}=\emptyset
\end{cases}.\label{e:Track_Distance}
\end{align}
Note that in \eqref{e:Track_Distance}, $\left\{ x\left(t\right)\right\} $
is a singleton if $t\in\mathcal{D}_{x}$, and $\left\{ x\left(t\right)\right\} =\emptyset$
if $t\notin\mathcal{D}_{x}$ (and likewise for $\left\{ y\left(t\right)\right\} $).
In this case, the OSPA distance defined in \eqref{e:OSPA_Distance}
reduces to
\begin{align}
d^{\left(c\right)}\left(\phi,\psi\right) & =\begin{cases}
0, & \left|\phi\right|=\left|\psi\right|=0\\
c, & \left|\phi\right|\neq\left|\psi\right|\\
\min\left(c,d\left(\phi,\psi\right)\right), & \left|\phi\right|=\left|\psi\right|=1
\end{cases}.\label{e:Track_Elem_Dist}
\end{align}
Note that the order parameter $p$ becomes redundant, and is therefore
omitted from \eqref{e:Track_Elem_Dist}.

In order to use \eqref{e:Track_Distance} as a base distance between
tracks, we need to establish that it defines a metric on $\mathbb{U}$.
That is, it must satisfy the properties P1-P4 as previously mentioned.
\begin{prop}
\label{p:Mean_OSPA_Metric}Let $d^{\left(c\right)}(\cdot,\cdot)$
be a metric on the finite subsets of $\mathbb{X}$, such that the
distance between a singleton and an empty set assumes the maximum
attainable value $c$. Then the distance between two tracks as defined
by \eqref{e:Track_Distance} is also a metric.
\end{prop}
The proof of this proposition is given in the appendix, which establishes
that \eqref{e:Track_Distance} is indeed a metric on the space $\mathbb{U}$,
when $d^{\left(c\right)}\left(\cdot,\cdot\right)$ is defined according
to \eqref{e:Track_Elem_Dist}. Before proceeding to define the OSPA\textsuperscript{(2)},
we make two important observations regarding the properties of this
base distance.
\begin{itemize}
\item Since $d^{\left(c\right)}\left(\cdot,\cdot\right)\leq c$, the distance
between tracks saturates at the value $c$, i.e. $\tilde{d}_{q}^{\left(c\right)}\left(\cdot,\cdot;w\right)\leq c$.
\item For two tracks $x$ and $y$ such that $\mathcal{D}_{x}=\mathcal{D}_{y}$,
\eqref{e:Track_Distance} can be interpreted as a mean square error
(MSE) between $x$ and $y$. Hence, the base distance can be regarded
as a generalisation of the MSE for tracks of different lengths and/or
domains.
\end{itemize}

\subsubsection{OSPA\protect\textsuperscript{(2)} for Tracks}

The distance between two tracks as defined in Section \ref{s:Track_Base_Distance}
is both a metric on the space $\mathbb{U}$, and bounded by the value
$c$. It is therefore suitable to serve as a base distance for the
OSPA metric on the space of finite sets of tracks $\mathcal{F}\left(\mathbb{U}\right)$.
Let $X=\left\{ x^{\left(1\right)},x^{\left(2\right)},\dots,x^{\left(m\right)}\right\} \subseteq\mathcal{F}\left(\mathbb{U}\right)$
and $Y=\left\{ y^{\left(1\right)},y^{\left(2\right)},\dots,y^{\left(n\right)}\right\} \subseteq\mathcal{F}\left(\mathbb{U}\right)$
be two sets of tracks, where $m\leq n$. We define the distance $\check{d}_{p}^{\left(c\right)}\left(X,Y\right)$
between $X$ and $Y$ as the OSPA with base distance $\tilde{d}^{\left(c\right)}\left(\cdot,\cdot\right)$
(the time averaged OSPA given by equation \eqref{e:Track_Distance}).
That is, 
\begin{align}
 & \check{d}_{p}^{\left(c\right)}\left(X,Y\right)\label{e:OSPA2_Distance}\\
 & =\left(\!\frac{1}{n}\!\left(\min_{\pi\in\Pi_{n}}\sum_{i=1}^{m}\tilde{d}^{\left(c\right)}\left(\!x^{\left(i\right)},y^{\left(\pi\left(i\right)\right)}\!\right)^{p}\!+\!c^{p}\left(n-m\right)\!\right)\!\right)^{1/p},\nonumber 
\end{align}
where $c$ is the cutoff and $p$ is the order parameter. Note that
if $m>n$, then $\check{d}_{p}^{\left(c\right)}\left(X,Y\right)\triangleq\check{d}_{p}^{\left(c\right)}\left(Y,X\right)$.
We refer to this as the OSPA\textsuperscript{(2)} distance, which
can be interpreted as the time-averaged per-track error.

\subsubsection{Efficient Evaluation of OSPA\protect\textsuperscript{(2)}}

Evaluating \eqref{e:OSPA2_Distance} involves the following three
steps:
\begin{enumerate}
\item Compute an $m\times n$ cost matrix $C$, where the $j$-th column
of the $i$-th row is given by $C_{i,j}=\tilde{d}_{q}^{\left(c\right)}\left(x^{\left(i\right)},y^{\left(j\right)}\right)$,
according to \eqref{e:Track_Distance}.
\item Use a 2-D optimal assignment algorithm on the matrix $C$, to find
the minimum cost 1-1 assignment of columns to rows.
\item Use the result of step 2 to evaluate $\check{d}_{p}^{\left(c\right)}\left(X,Y\right)$
according to \eqref{e:OSPA2_Distance}.
\end{enumerate}
A basic implementation of this procedure would require computing the
base distance between all pairs of tracks in $X$ and $Y$, which
has complexity $\mathcal{O}\left(\left|w_{k}\right|mn\right)$. Step
2 then requires solving a dense optimal assignment problem with complexity
$\mathcal{O}\left(\left(m+n\right)^{3}\right)$. This would preclude
its use in large-scale tracking scenarios involving millions of objects,
as the cost matrix would consume too much memory, and the assignment
problem would infeasible to solve.

Fortunately, in many practical applications, the base distance between
most pairs of tracks will saturate at the cutoff value $c$. This
can be exploited to dramatically reduce the computational complexity,
making it feasible for large-scale problems. Firstly, recall that
the time averaging in the base distance \eqref{e:Track_Distance}
is carried out only over the union of the track domains. Consequently,
the base distance between any pair of tracks that have no corresponding
states closer than a distance of $c$, must saturate at $c$. Such
pairs can be considered unassignable, and efficient spatial searching
algorithms can be applied to extract only the assignable pairs of
tracks in $\mathcal{O}\left(\left|w_{k}\right|m\log n\right)$ time.
Once these have been found, a sparse optimal assignment algorithm
can be used to obtain the lowest-cost matching between the ground
truth and estimated trajectories. Such algorithms can solve sparse
assignment problems with a much lower average complexity than is possible
under the dense optimal assignment formulation.

\subsubsection{Visualisation of OSPA\protect\textsuperscript{(2)}}

In practice, it is desirable to examine the tracking performance as
a function of time, so that trends in algorithm behaviour can be analysed
in response to changing scenario conditions. This can be achieved
by plotting 
\begin{align}
\alpha_{k}\left(X,Y;w_{k}\right) & =\check{d}_{p}^{\left(c\right)}\left(X_{w_{k}},Y_{w_{k}}\right)
\end{align}
as a function of $k$, where $w_{k}$ is a set of time indices that
varies with $k$, and
\begin{align}
X_{w_{k}} & =\left\{ x\vert_{w_{k}}:x\in X\text{ and }\mathcal{D}_{x}\cap w_{k}\neq\emptyset\right\} ,\\
Y_{w_{k}} & =\left\{ y\vert_{w_{k}}:y\in Y\text{ and }\mathcal{D}_{y}\cap w_{k}\neq\emptyset\right\} ,
\end{align}
where $f\vert_{w}$ denotes the restriction of $f$ to domain $w$. 

Note that the sets $X_{w_{k}}$ and $Y_{w_{k}}$ only contain those
tracks whose domain overlaps with $w_{k}$, i.e. any tracks whose
domain lies completely outside the set $w_{k}$ are disregarded. Choosing
different values for the set $w_{k}$ allows us to examine the performance
of tracking algorithms over different time scales. For example, a
straightforward approach is to set $w_{k}=\left\{ k-N+1,k-N+2,\dots,k\right\} $,
so that at time $k$, the set $w_{k}$ consists of only the latest
$N$ time steps. In this case, choosing a small value for $N$ will
indicate the tracking performance over shorter time periods, while
larger values will reveal the longer-term tracking performance. This
choice is highly dependent on the application at hand. For example,
in real-time surveillance, we may only be interested in tracking objects
over a period of a few minutes, as older information may be considered
irrelevant. In this case, a small value for $N$ would suffice to
capture the important aspects of the tracking performance. However,
in an off-line scenario analysis application, we might require accurate
trajectory estimates over much longer time periods, in which case
a larger value for $N$ would be more appropriate.
\begin{rem}
Note that computing the OSPA\textsuperscript{(2)} (for sets of tracks)
on a sliding window as described above, converges to the traditional
OSPA (for sets of points) as $N$ becomes smaller. For $N=1$ the
OSPA\textsuperscript{(2)} becomes identical to the traditional OSPA.
\end{rem}
It is important to understand that the OSPA\textsuperscript{(2)}
distance has a different interpretation to that of the traditional
OSPA distance. Whereas the traditional OSPA distance captures the
error between the true and estimated multi-target states at a single
instant in time, the OSPA\textsuperscript{(2)} distance captures
the error between the true and estimated sets of tracks over a range
of time instants, as determined by the choice of $w_{k}$. Therefore,
careful consideration must be given to the design of $w_{k}$, and
the user must be mindful of this when interpreting the results.\clearpage{}

\subsection{Numerical Results}

In this section, we demonstrate the scalability of the proposed GLMB
filter implementation, by applying it to a simulated multi-target
tracking scenario where the peak number of objects appearing simultaneously
exceeds one million. The scenario runs for $1000$ time steps, and
the surveillance region is a 2D area with width $16$ km and height
$9$ km. New objects are born during the time intervals $\left[0,400\right]$
and $\left[600,800\right]$. The positions of new targets are simulated
by drawing random samples from a Gaussian mixture with $2000$ equally
weighted components, where the means are distributed uniformly over
the 2D region $\left[1,15\right]\times\left[1,8\right]$ km, and the
covariances are inverse-Wishart distributed with scale matrix $1000\times I_{2}$
and $4$ degrees of freedom. The velocities of new targets are simulated
by sampling their courses uniformly on the interval $[0,2\pi]$ radians,
and their speeds uniformly on the interval $[1,8]$ metres per second.
The rate of target birth varies with time, with the highest rate occurring
in the time intervals $\left[1,100\right]$ and $\left[200,300\right]$.

The peak cardinality of $1,082,974$ objects occurs at time $400$.
At this time, the mean target density is around $7,500$ objects per
square kilometre, however, since the objects are not uniformly distributed
in space, the peak density is approximately $32,700$ objects per
square kilometre. The survival time for each object is chosen uniformly
at random in the interval $\left[400,700\right]$, and the tracker
assumes a survival probability of $0.999$. The measurement noise
standard deviation is $0.15$ metres in both the x and y directions,
and the detection probability is constant at $0.9$. The false alarms
are uniformly distributed across the space, with a Poisson cardinality
distribution with a mean of $250,000$ per scan. In the label partitioning
step within the tracker, the maximum number of labels per group is
set to $20$.

The tracking algorithm was coded in C++, making use of OpenMP to implement
parallel processing wherever possible. We executed the algorithm on
a machine with four 16-core AMD Opteron 6376 processors (for a total
of 64 physical processor cores), and 256 GB of memory. On this hardware
configuration, the peak time taken to process a frame of measurements
was approximately 5 minutes when the cardinality was highest, but
the algorithm ran considerably faster than this at times when there
were fewer objects in the scene. The peak memory usage of the algorithm
was approximately 50 GB. To evaluate the tracking performance, the
OSPA\textsuperscript{(2)} metric was coded in Matlab, and we used
the ``parfor'' construct to parallelise some aspects of the computation.
The average time taken to evaluate each point in the OSPA\textsuperscript{(2)}
curve was approximately 1 minute.

\begin{figure}
\begin{centering}
\includegraphics[width=1\columnwidth]{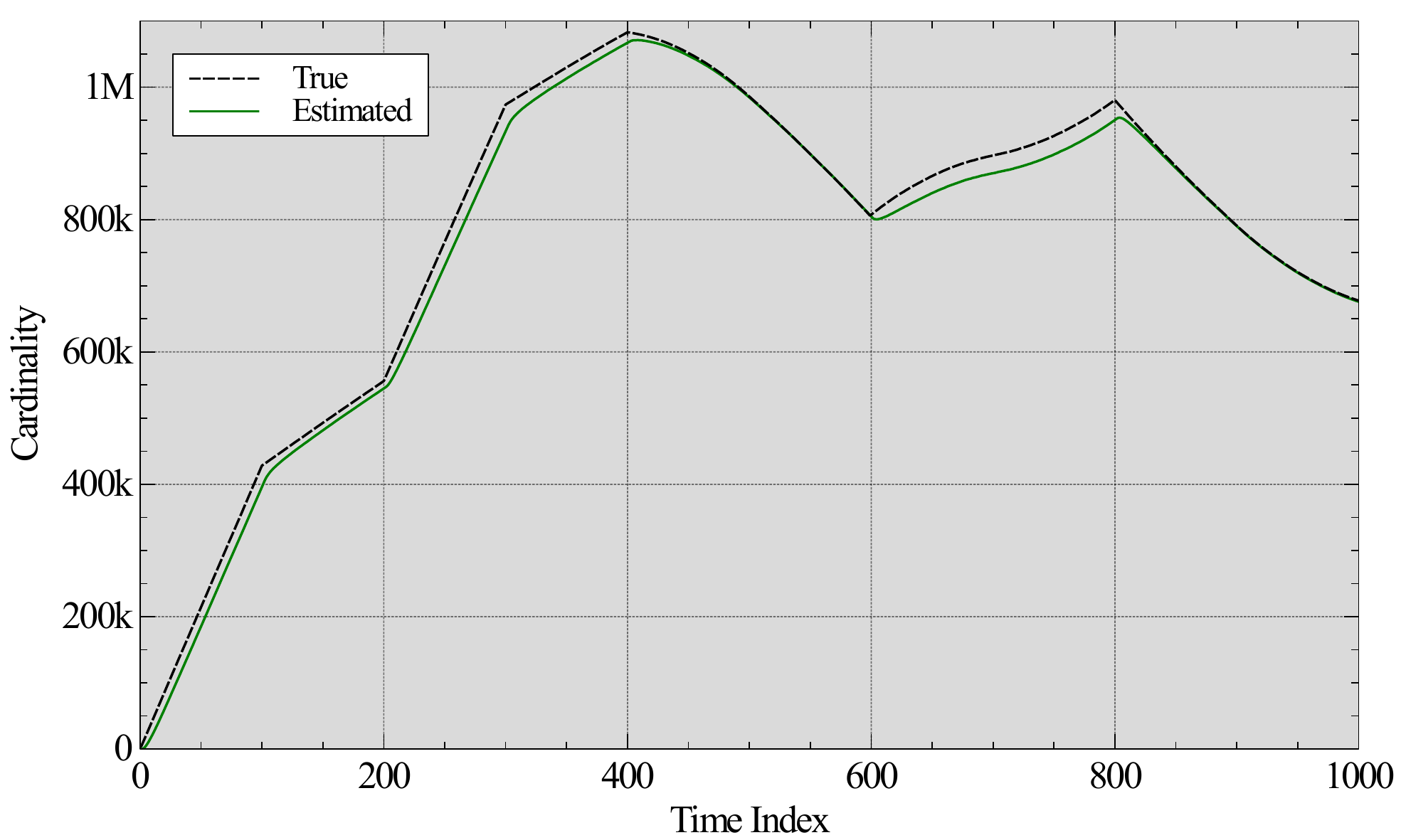}
\par\end{centering}
\caption{True and estimated cardinality for large-scale tracking scenario}

\label{f:Cardinality_1M}
\end{figure}

\begin{figure}
\begin{centering}
\includegraphics[width=1\columnwidth]{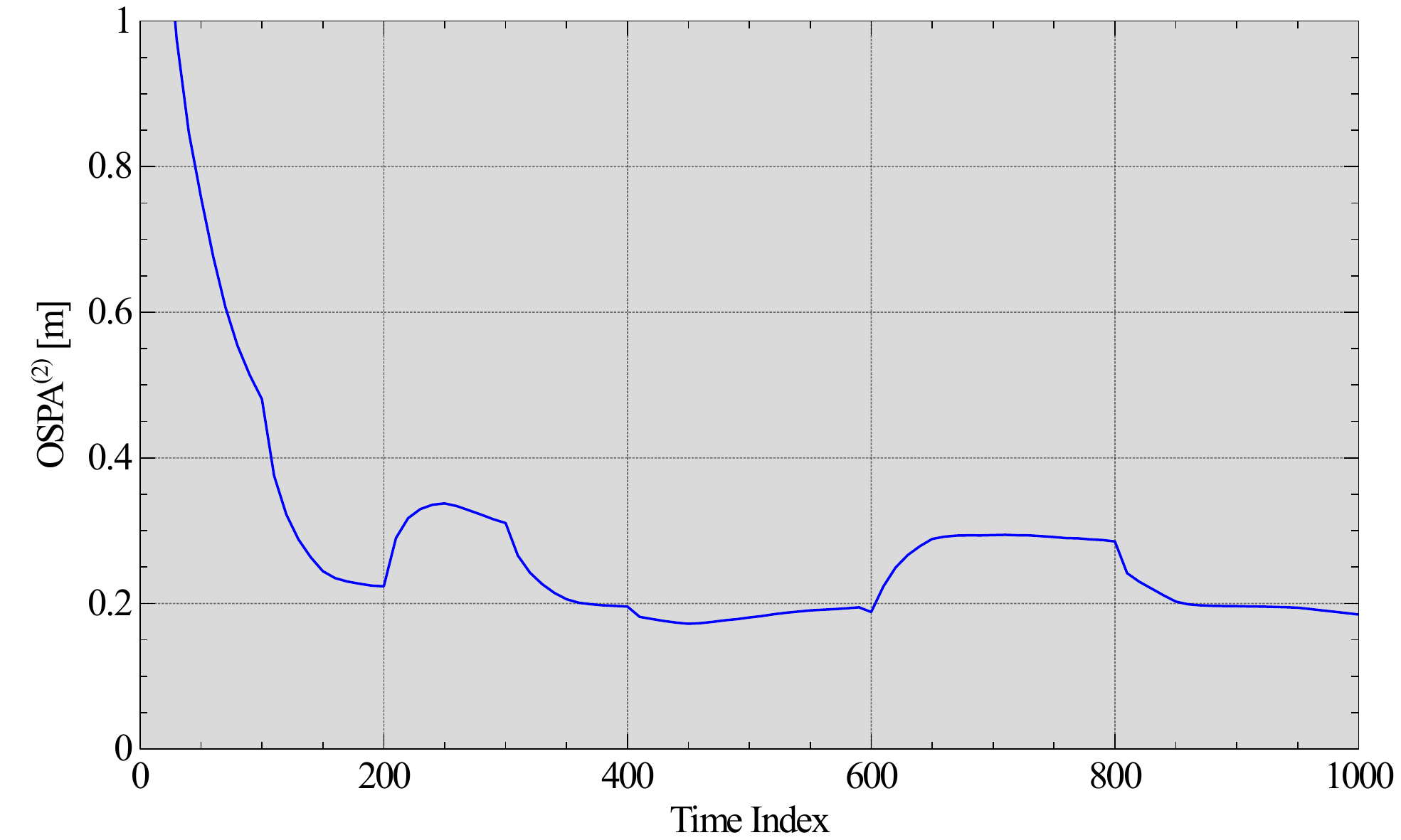}
\par\end{centering}
\caption{OSPA\protect\textsuperscript{(2)} distance for large-scale tracking
scenario}

\label{f:OSPA2_1M}
\end{figure}

\begin{figure*}
\begin{centering}
\includegraphics[width=1\textwidth]{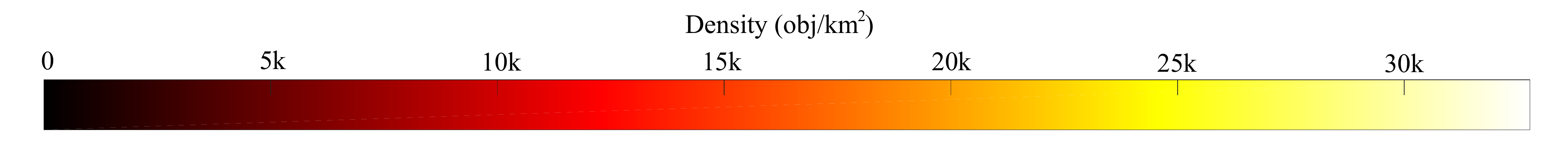}
\par\end{centering}
\begin{centering}
\subfloat[Time 50]{\begin{centering}
\includegraphics[width=0.5\textwidth]{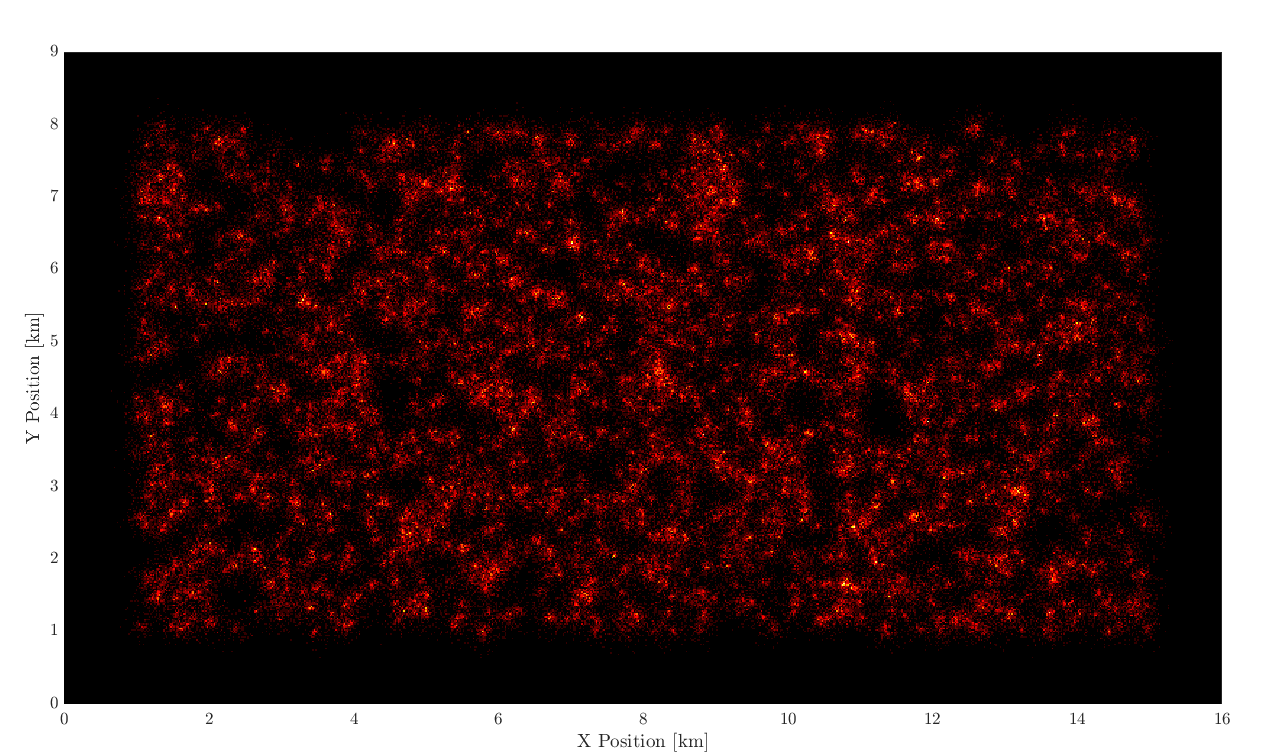}\includegraphics[width=0.5\textwidth]{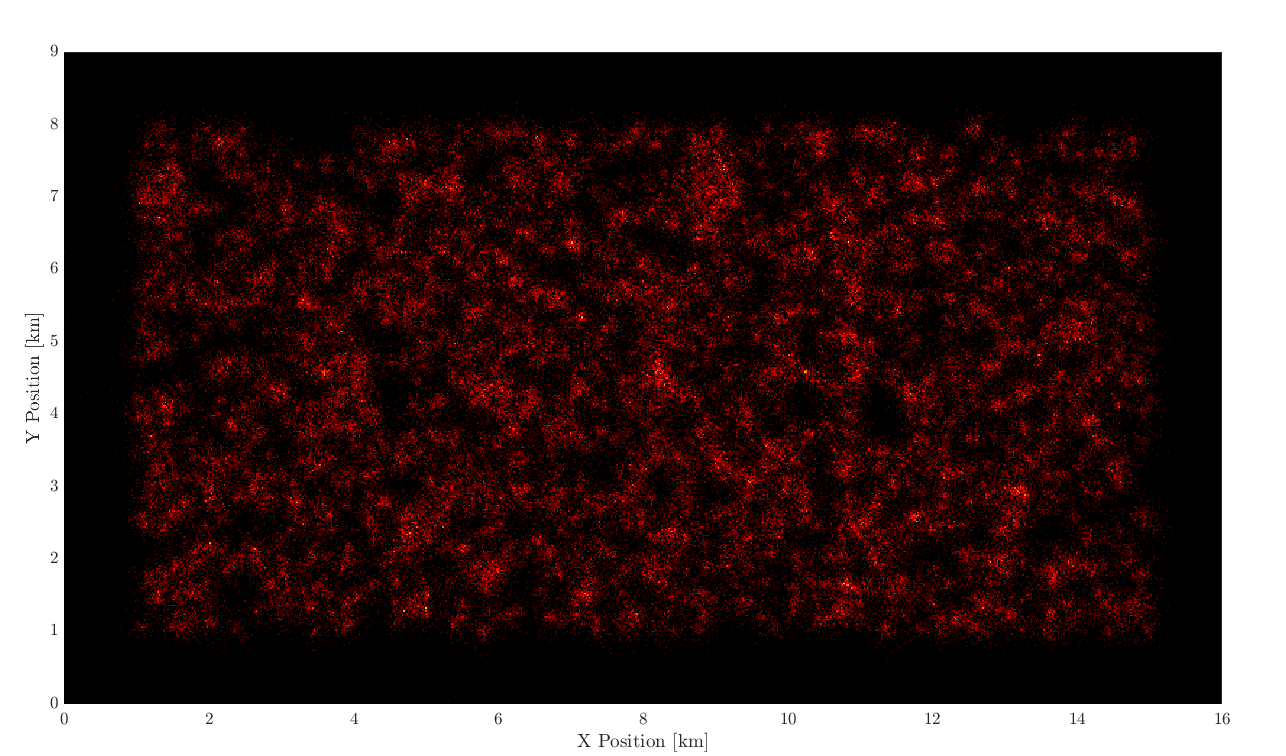}
\par\end{centering}

}
\par\end{centering}
\begin{centering}
\subfloat[Time 200]{\begin{centering}
\includegraphics[width=0.5\textwidth]{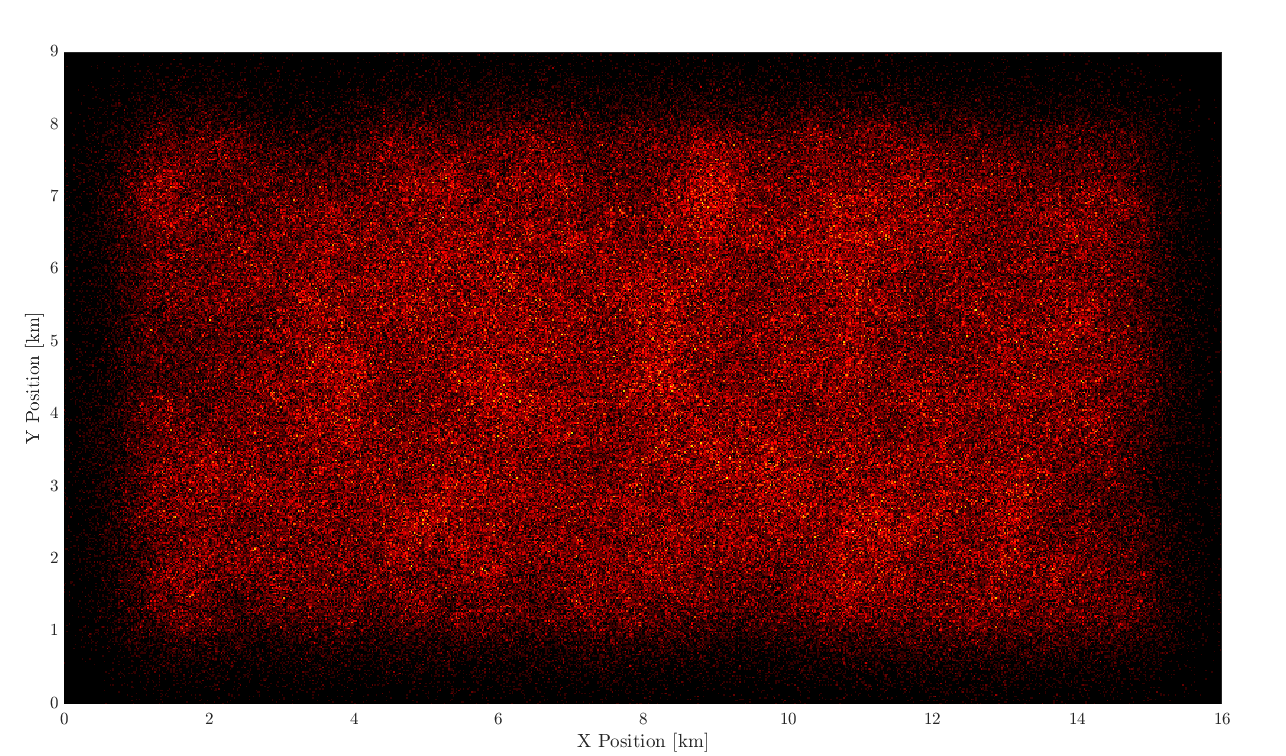}\includegraphics[width=0.5\textwidth]{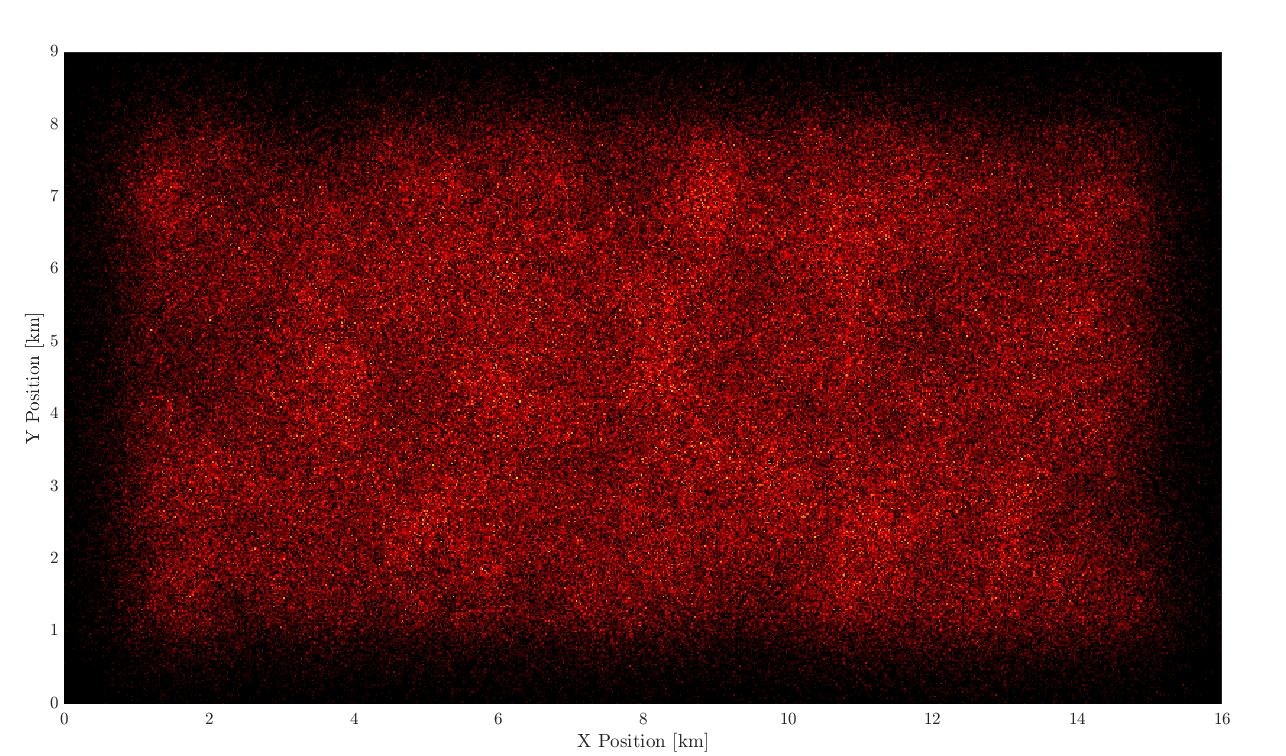}
\par\end{centering}

}
\par\end{centering}
\begin{centering}
\subfloat[Time 400]{\begin{centering}
\includegraphics[width=0.5\textwidth]{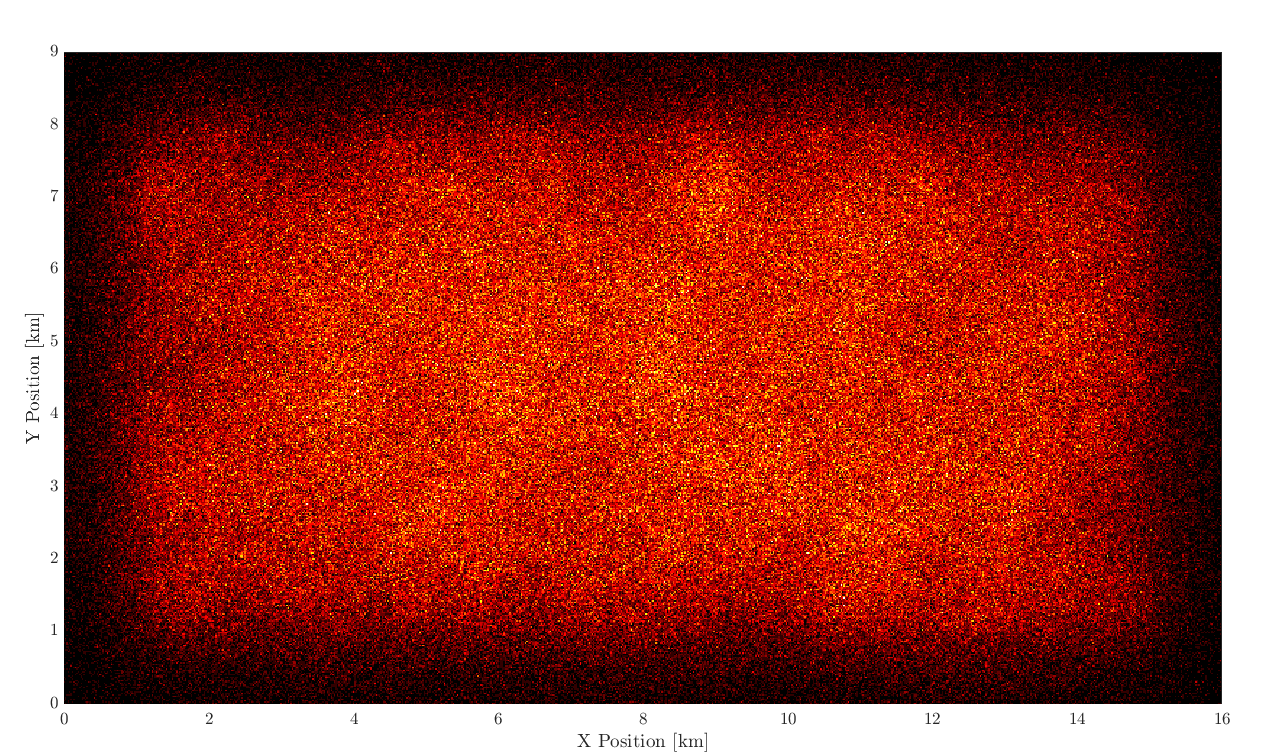}\includegraphics[width=0.5\textwidth]{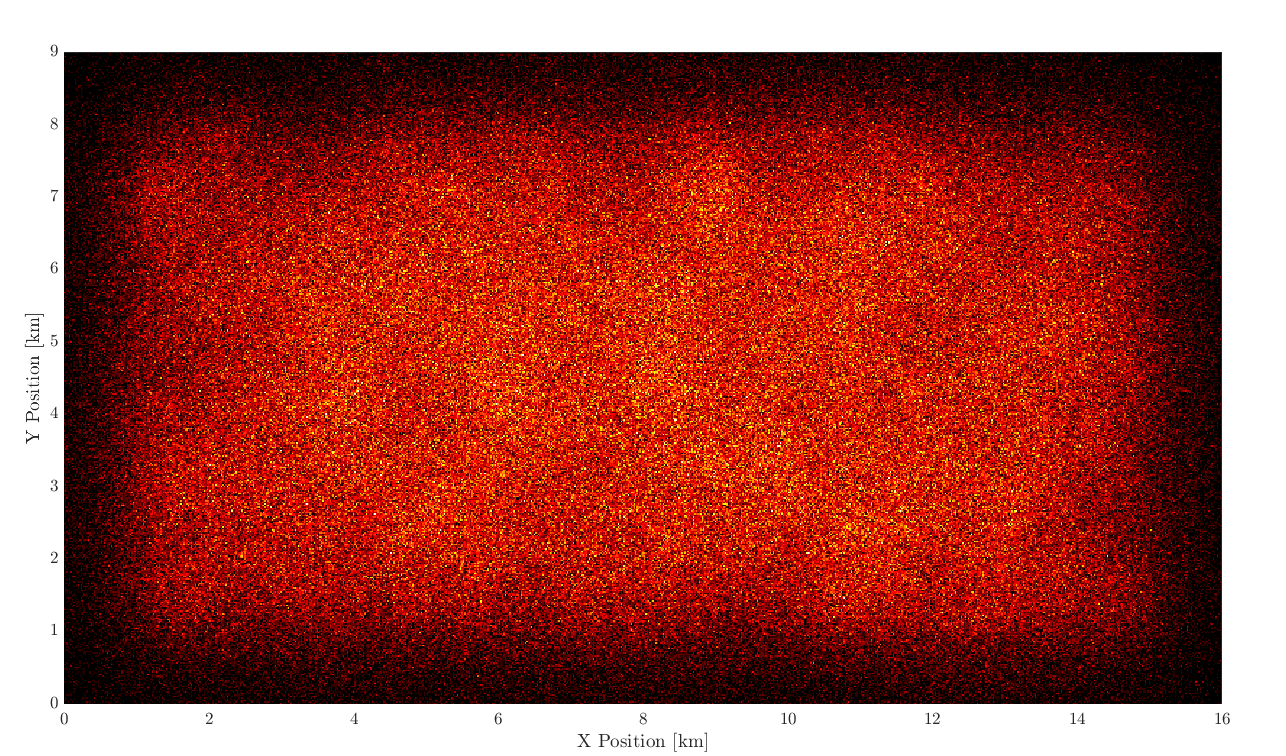}
\par\end{centering}

}
\par\end{centering}
\caption{True (left) and estimated (right) target density at times 50, 200
and 400}

\label{f:Target_Density_1}
\end{figure*}

\begin{figure*}
\begin{centering}
\includegraphics[width=1\textwidth]{figures/colorbar}
\par\end{centering}
\begin{centering}
\subfloat[Time 600]{\begin{centering}
\includegraphics[width=0.5\textwidth]{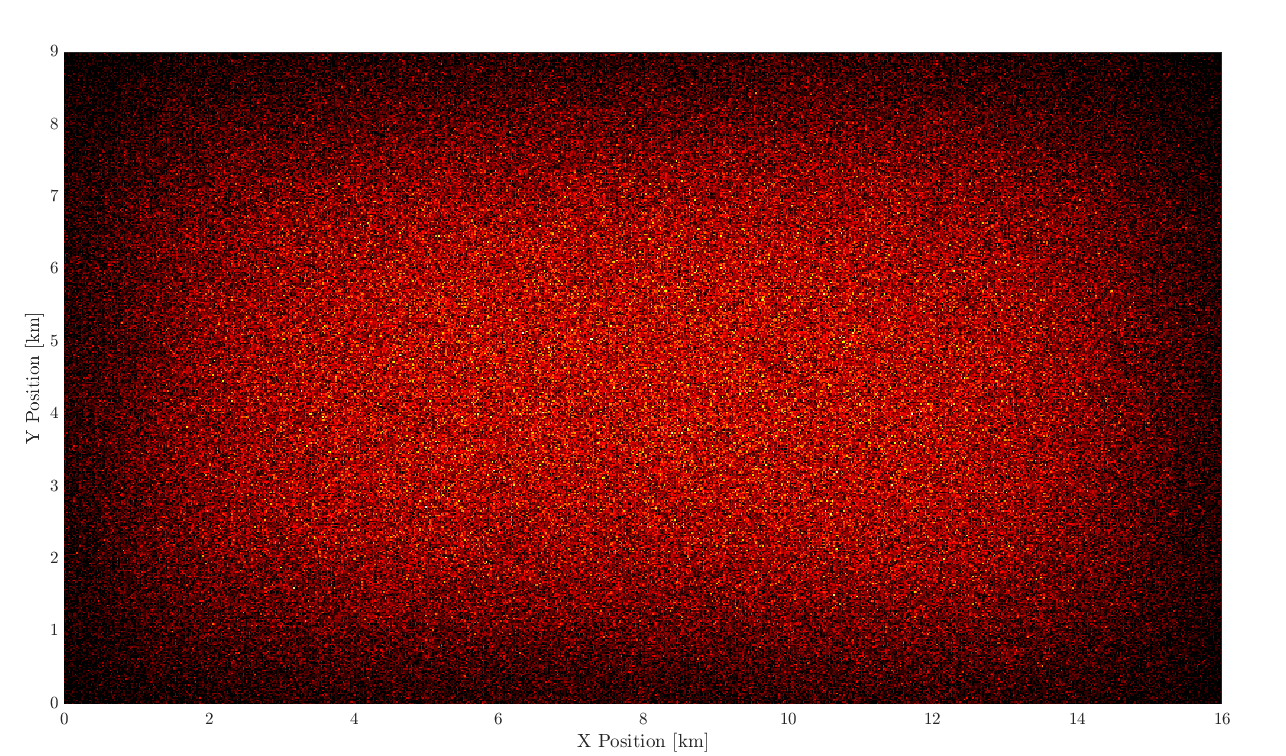}\includegraphics[width=0.5\textwidth]{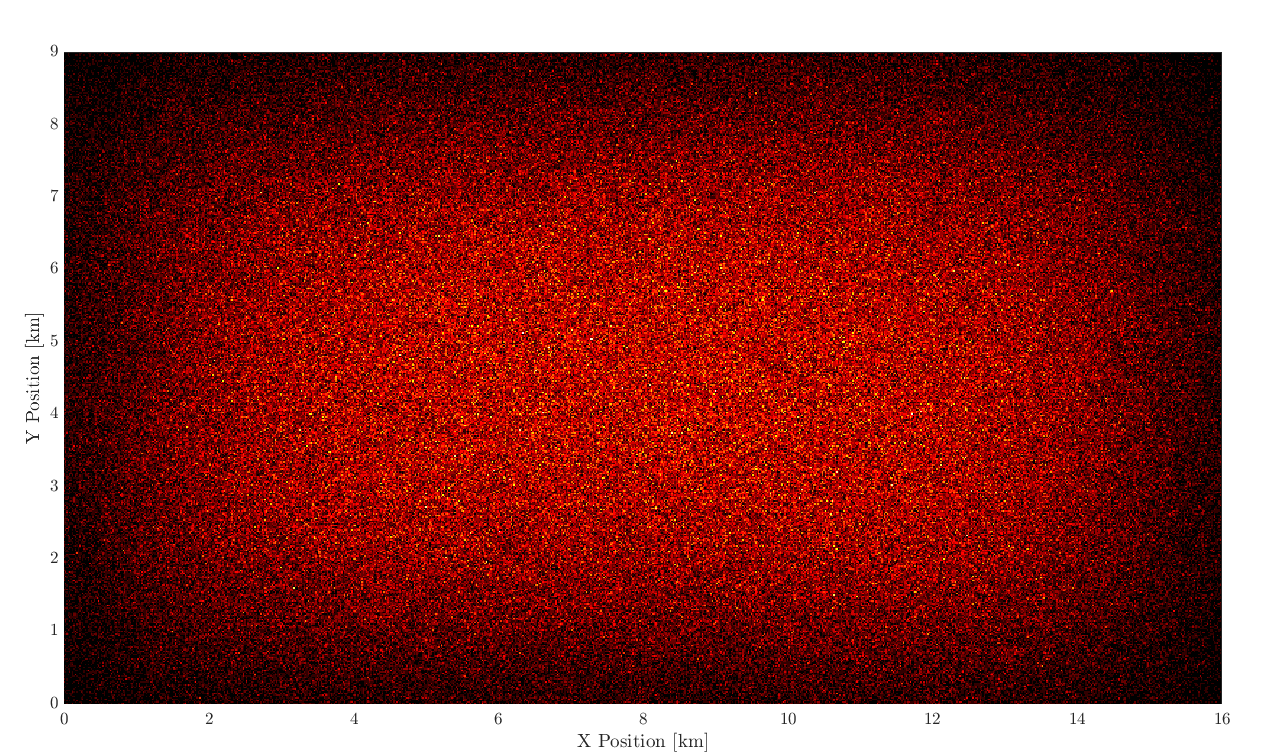}
\par\end{centering}

}
\par\end{centering}
\begin{centering}
\subfloat[Time 800]{\begin{centering}
\includegraphics[width=0.5\textwidth]{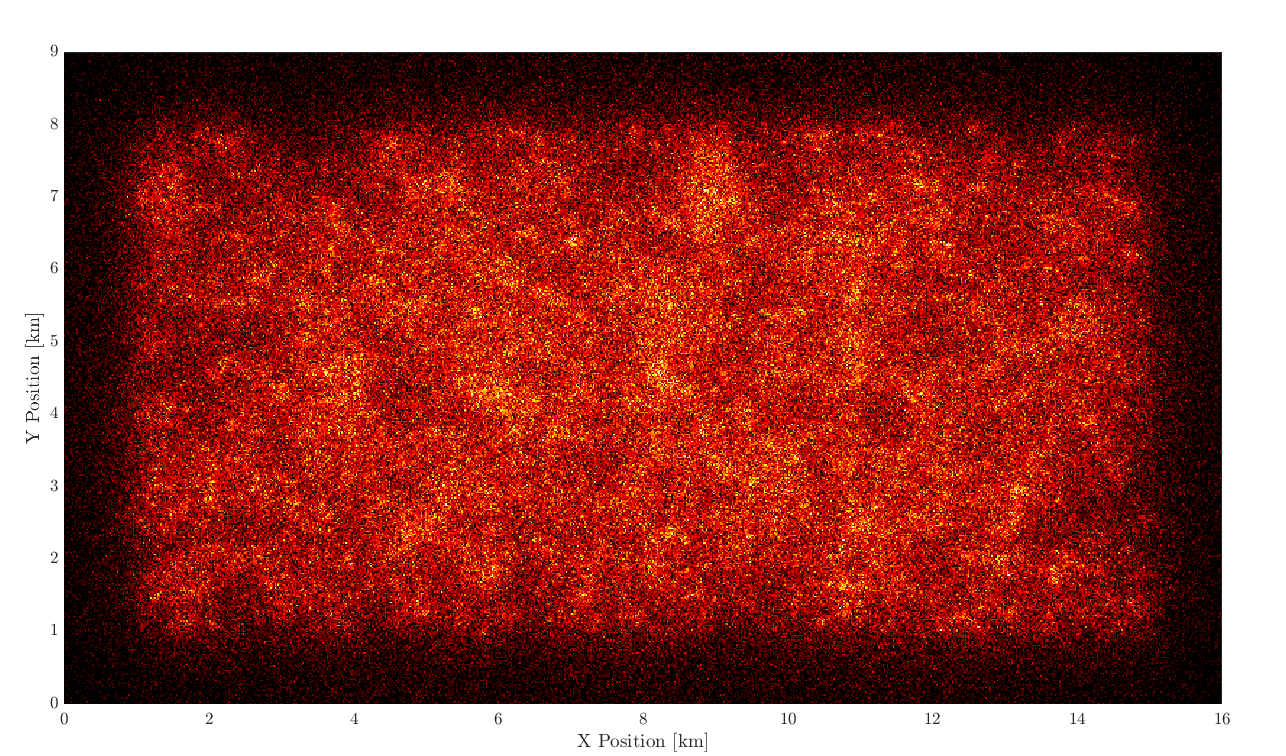}\includegraphics[width=0.5\textwidth]{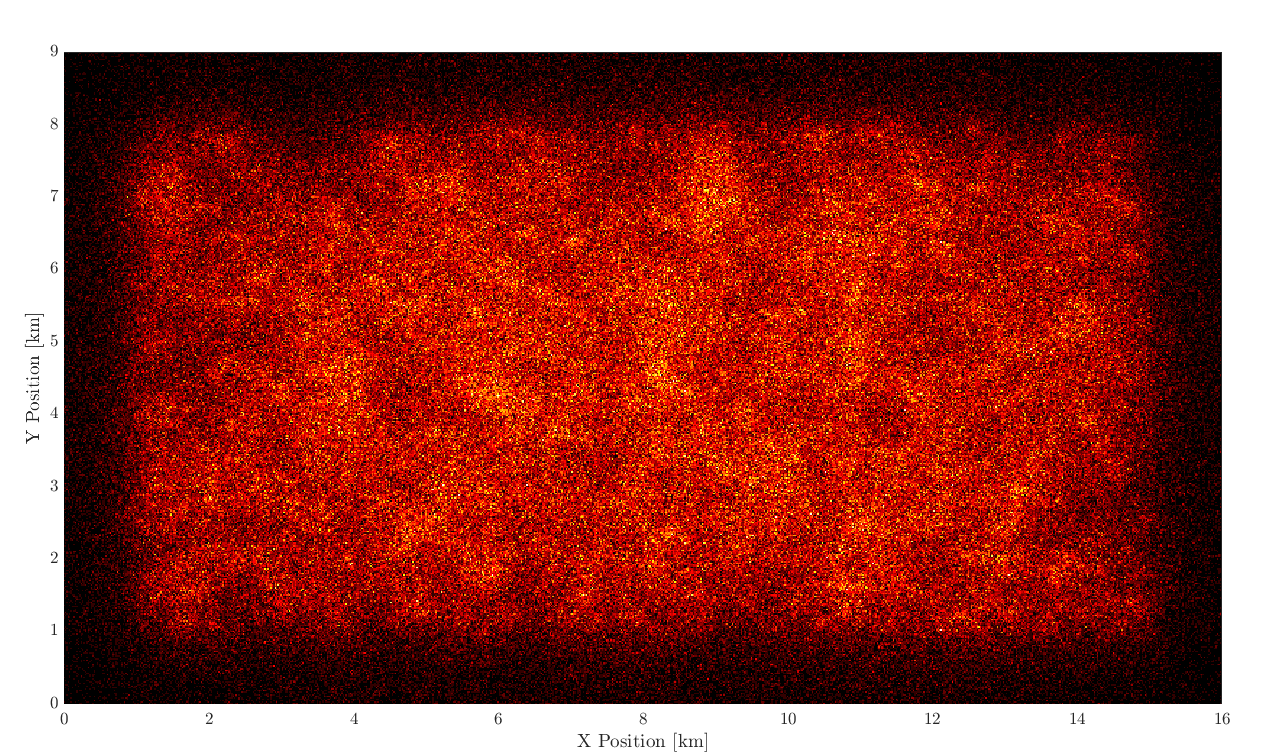}
\par\end{centering}

}
\par\end{centering}
\begin{centering}
\subfloat[Time 1000]{\begin{centering}
\includegraphics[width=0.5\textwidth]{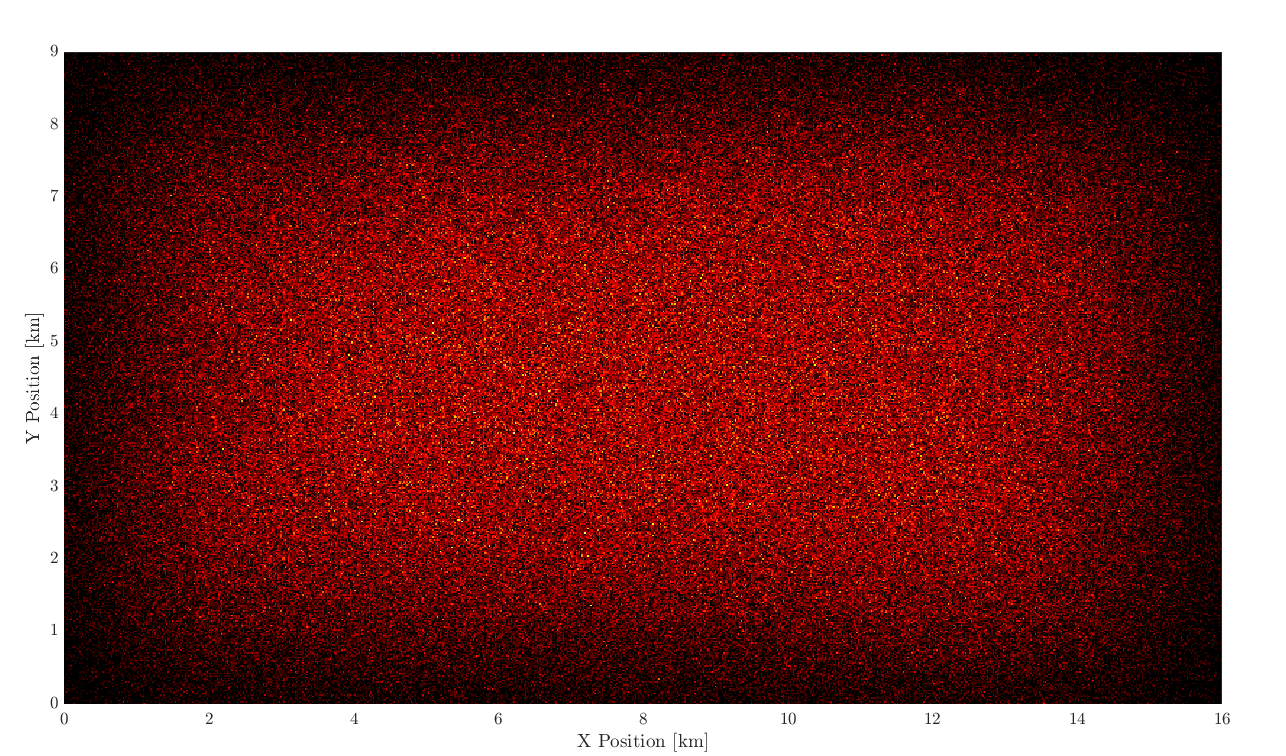}\includegraphics[width=0.5\textwidth]{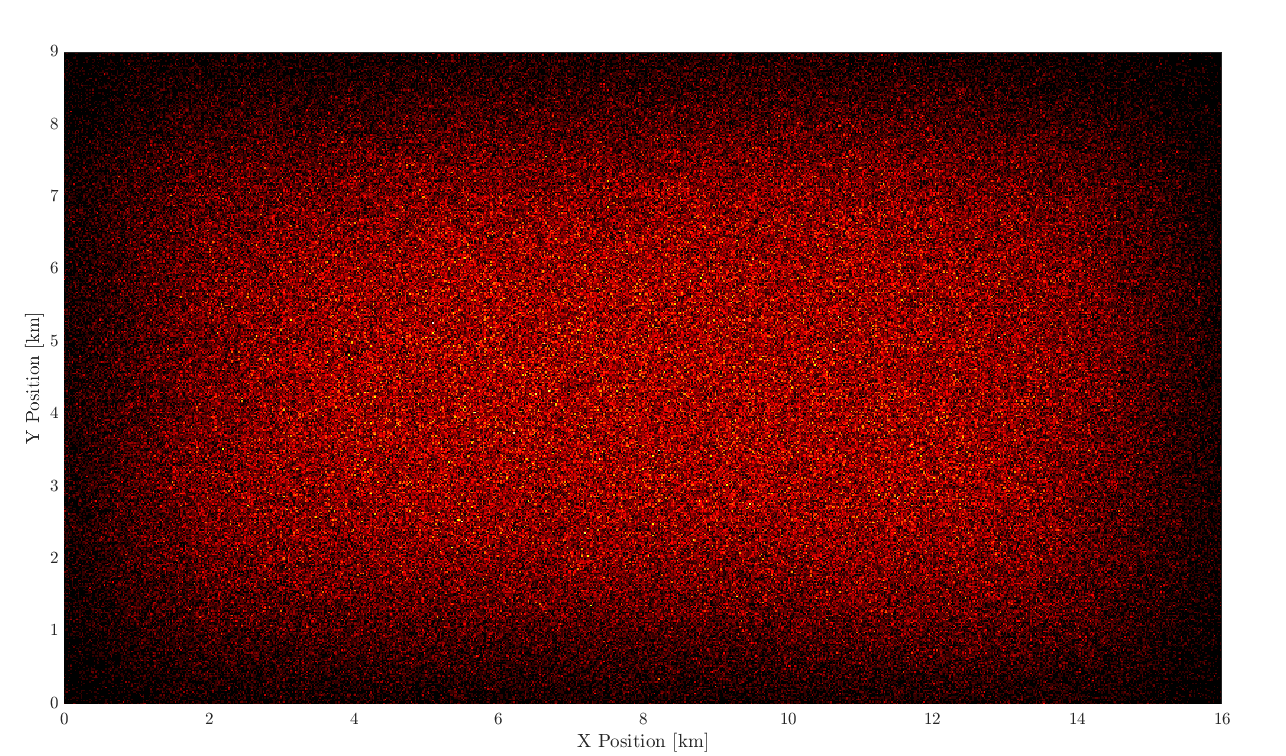}
\par\end{centering}

}
\par\end{centering}
\caption{True (left) and estimated (right) target density at times 600, 800
and 1000}

\label{f:Target_Density_2}
\end{figure*}

For a scenario of this scale, it is not practical to show the trajectories
of individual objects, thus we resort to showing summary statistics.
The true and estimated cardinality is shown in Figure \ref{f:Cardinality_1M},
and the OSPA\textsuperscript{(2)} distance is shown in Figure \ref{f:OSPA2_1M}.
Figures \ref{f:Target_Density_1} and \ref{f:Target_Density_2} contain
six snapshots of the true and estimated target density at various
times during the scenario. For the OSPA\textsuperscript{(2)} calculation,
the cutoff was set to $c=2$, the order was $p=1$, and a sliding
window over the latest $50$ time steps was used to evaluate each
point in the curve.

From the cardinality plot, it can be seen that the estimated cardinality
lags behind the true cardinality during the times when new targets
are being born. This is to be expected, as the measurement-driven
birth model needs to consider a very large number of potential birth
tracks at each scan. To avoid initiating too many false tracks, the
filter delays the initiation of tracks until there is more data to
confirm the presence of an object. The OSPA\textsuperscript{(2)}
plot shows increased error during the periods in which new targets
are appearing, due the delay in initiating new tracks. At other times,
the error stabilises to approximately $0.2$ metres per object per
unit time.

\section{Conclusion}

We have presented an efficient and scalable implementation of the
generalised labeled multi-Bernoulli filter, that is capable of estimating
the trajectories of a very large number of objects simultaneously,
in the order of millions per frame. The proposed method makes efficient
use of the available computational resources, by decomposing large-scale
tracking problems into smaller independent sub-problems. The decomposition
is carried out via marginalisation of high-dimensional multi-object
densities, using a technique that is shown to be optimal in the sense
that it minimises the KLD for a given partition of the object label
space. This allows the algorithm to fully exploit the potential of
highly parallel processing, as afforded by modern multi-core computing
architectures. Due to its relatively low processing time and memory
requirements, simulations show that the proposed technique is capable
of tracking in excess of a million objects simultaneously, running
on standard off-the-shelf computing equipment. Additionally, we have
introduced a new way of using the OSPA metric, that captures more
information about the multi-object tracking performance. We have demonstrated
that this metric can be feasibly evaluated in large-scale tracking
scenarios.

\appendix{}
\begin{center}
\textbf{Proof of Proposition \ref{p:Mean_OSPA_Metric}}
\par\end{center}

Firstly, since $d^{\left(c\right)}\left(\cdot,\cdot\right)\geq0$
and $\left|\mathcal{D}_{x}\cup\mathcal{D}_{y}\right|\geq0$, all terms
in the summation over $t\in\mathcal{D}_{x}\cup\mathcal{D}_{y}$ are
clearly non-negative. Hence $\tilde{d}^{\left(c\right)}\left(\cdot,\cdot\right)$
satisfies metric property P1.

Second, $\tilde{d}^{\left(c\right)}\left(x,y\right)=0$ if an only
if $x=y=\emptyset$, or $d^{\left(c\right)}\left(\left\{ x\left(t\right)\right\} ,\left\{ y\left(t\right)\right\} \right)=0$
for all $t\in\mathcal{D}_{x}\cup\mathcal{D}_{y}$. Since $d^{\left(c\right)}\left(\cdot,\cdot\right)$
is a metric, $d^{\left(c\right)}\left(\left\{ x\left(t\right)\right\} ,\left\{ y\left(t\right)\right\} \right)=0\iff\left\{ x\left(t\right)\right\} =\left\{ y\left(t\right)\right\} $.
Hence $\tilde{d}^{\left(c\right)}\left(x,y\right)=0\iff x=y$, satisfying
metric property P2.

Third, since $d^{\left(c\right)}\left(\cdot,\cdot\right)$ is a metric,
and $\mathcal{D}_{x}\cup\mathcal{D}_{y}=\mathcal{D}_{y}\cup\mathcal{D}_{x}$,
all terms in \eqref{e:Track_Distance} are symmetric in their arguments.
Hence $\tilde{d}^{\left(c\right)}\left(\cdot,\cdot\right)$ satisfies
metric property P3.

It remains to verify metric property P4, namely the triangle inequality,
which is accomplished via induction. Since $d^{\left(c\right)}\left(\cdot,\cdot\right)$
is a metric, it is straightforward to show that the distance between
the tracks at a single time instant $t_{1}$ satisfies the triangle
inequality. Let us assume that the distance between the tracks over
time instants $t_{1},t_{2},\dots,t_{k}$ satisfies the triangle inequality.
We now proceed to show that the distance between the tracks over time
instants $t_{1},t_{2},\dots,t_{k},t_{k+1}$ also satisfies the triangle
inequality.

\onecolumn

When at least one of the sets $\mathcal{D}_{x}\cup\mathcal{D}_{y}$,
$\mathcal{D}_{y}\cup\mathcal{D}_{z}$, $\mathcal{D}_{z}\cup\mathcal{D}_{x}$
is empty, the triangle inequality for tracks over time instants $t_{1},t_{2},\dots,t_{k},t_{k+1}$
can be easily verified, since $d^{\left(c\right)}\left(\cdot,\cdot\right)$
is a metric. Hence, we consider the case where $\mathcal{D}_{x}\cup\mathcal{D}_{y}$,
$\mathcal{D}_{y}\cup\mathcal{D}_{z}$, $\mathcal{D}_{z}\cup\mathcal{D}_{x}$
are all non-empty.

Before proceeding with proof, we must define some notation. Let us
denote the cardinalities of the basic sets in $\mathcal{D}_{x}\cup\mathcal{D}_{y}\cup\mathcal{D}_{z}$
by
\begin{align}
N & \triangleq\left\vert \mathcal{D}_{x}\cap\mathcal{D}_{y}\cap\mathcal{D}_{z}\right\vert ,\label{e:N}\\
N_{\breve{x}} & \triangleq\left\vert \mathcal{D}_{x}-\mathcal{D}_{y}-\mathcal{D}_{z}\right\vert ,\quad N_{p}\triangleq\left\vert \mathcal{D}_{x}\cap\mathcal{D}_{y}-\mathcal{D}_{z}\right\vert ,\label{e:NxNa}\\
N_{\breve{y}} & \triangleq\left\vert \mathcal{D}_{y}-\mathcal{D}_{z}-\mathcal{D}_{x}\right\vert ,\quad N_{q}\triangleq\left\vert \mathcal{D}_{y}\cap\mathcal{D}_{z}-\mathcal{D}_{x}\right\vert ,\label{e:NyNb}\\
N_{\breve{z}} & \triangleq\left\vert \mathcal{D}_{z}-\mathcal{D}_{x}-\mathcal{D}_{y}\right\vert ,\quad N_{r}\triangleq\left\vert \mathcal{D}_{z}\cap\mathcal{D}_{x}-\mathcal{D}_{y}\right\vert ,\label{e:NzNc}
\end{align}
as illustrated in the diagram below.
\begin{center}
\includegraphics[width=0.4\columnwidth]{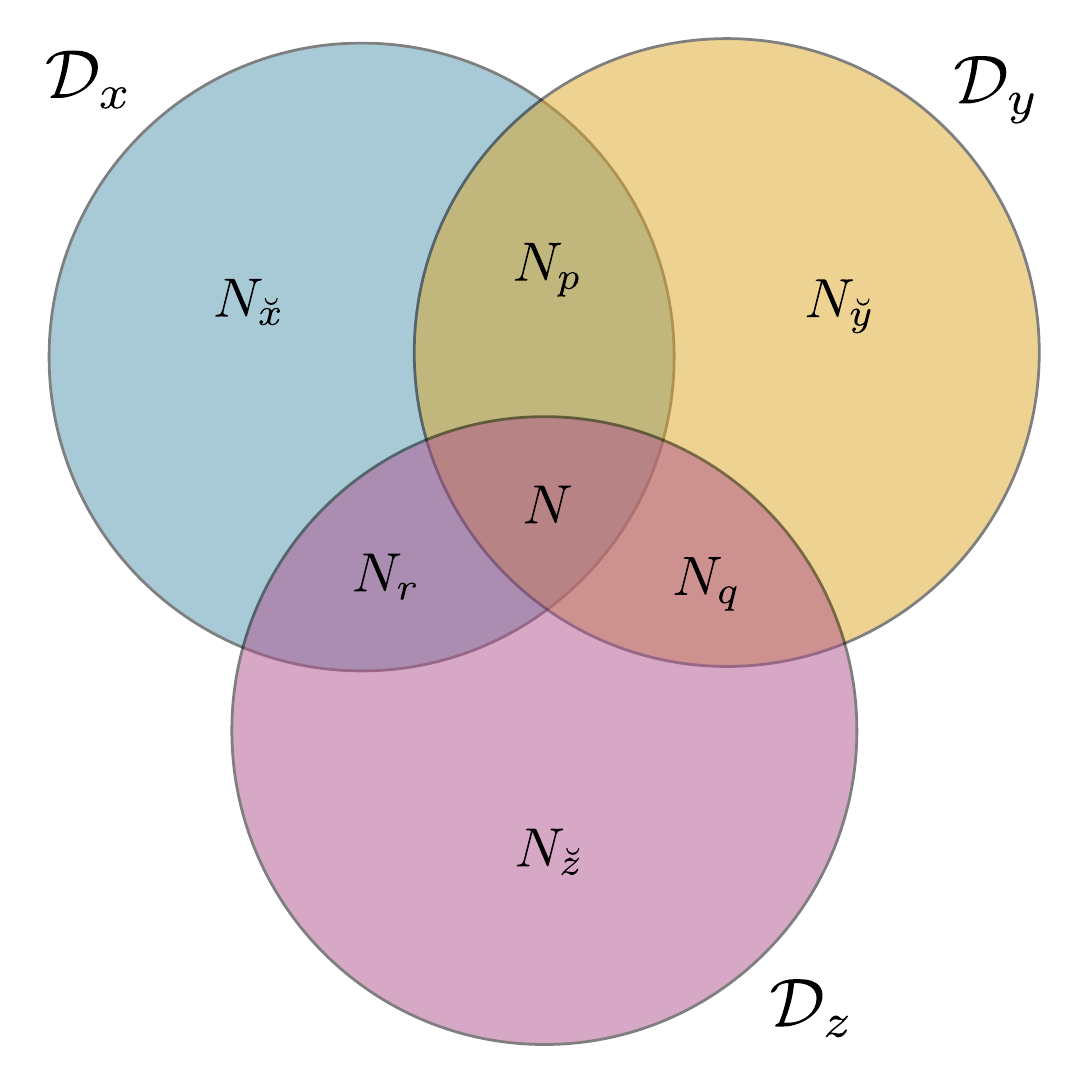}
\par\end{center}

\begin{flushleft}
Furthermore, we adopt the following abbreviations:
\begin{align}
S & \triangleq N+N_{p}+N_{q}+N_{r},\label{e:S}\\
T & \triangleq S+\left\vert \mathcal{D}_{x}\cup\mathcal{D}_{y}\cup\mathcal{D}_{z}\right\vert =S+S+N_{\breve{x}}+N_{\breve{y}}+N_{\breve{z}}=2S+N_{\breve{x}}+N_{\breve{y}}+N_{\breve{z}},\label{e:T}\\
P & \triangleq\left\vert \mathcal{D}_{x}\cup\mathcal{D}_{y}\right\vert =N+N_{p}+N_{q}+N_{r}+N_{\breve{x}}+N_{\breve{y}}=S+N_{\breve{x}}+N_{\breve{y}},\label{e:P}\\
Q & \triangleq\left\vert \mathcal{D}_{y}\cup\mathcal{D}_{z}\right\vert =N+N_{p}+N_{q}+N_{r}+N_{\breve{y}}+N_{\breve{z}}=S+N_{\breve{y}}+N_{\breve{z}},\label{e:Q}\\
R & \triangleq\left\vert \mathcal{D}_{z}\cup\mathcal{D}_{x}\right\vert =N+N_{p}+N_{q}+N_{r}+N_{\breve{x}}+N_{\breve{z}}=S+N_{\breve{x}}+N_{\breve{z}},\label{e:R}\\
p & \triangleq{\textstyle \sum\limits _{t\in\mathcal{D}_{x}\cup\mathcal{D}_{y}}}d\left(\left\{ x\left(t\right)\right\} ,\left\{ y\left(t\right)\right\} \right),\text{ \ \ \ }p^{\prime}\triangleq d^{\left(c\right)}\left(\left\{ x\left(k+1\right)\right\} ,\left\{ y\left(k+1\right)\right\} \right),\label{e:p_p'}\\
q & \triangleq{\textstyle \sum\limits _{t\in\mathcal{D}_{y}\cup\mathcal{D}_{z}}}d\left(\left\{ y\left(t\right)\right\} ,\left\{ z\left(t\right)\right\} \right),\text{ \ \ \ }q^{\prime}\triangleq d^{\left(c\right)}\left(\left\{ y\left(k+1\right)\right\} ,\left\{ z\left(k+1\right)\right\} \right),\label{e:q_q'}\\
r & \triangleq{\textstyle \sum\limits _{t\in\mathcal{D}_{z}\cup\mathcal{D}_{x}}}d\left(\left\{ z\left(t\right)\right\} ,\left\{ x\left(t\right)\right\} \right),\text{ \ \ \ }r^{\prime}\triangleq d^{\left(c\right)}\left(\left\{ z\left(k+1\right)\right\} ,\left\{ x\left(k+1\right)\right\} \right).\label{e:r_r'}
\end{align}
The sum $p$ over (the non-empty) $\mathcal{D}_{x}\cup\mathcal{D}_{y}$
can be decomposed into; $\hat{p}_{N}$ consisting of $N$ terms on
$\left(\mathcal{D}_{x}\cap\mathcal{D}_{y}\cap\mathcal{D}_{z}\right)$,
$\hat{p}_{N_{p}}$ consisting of $N_{p}$ terms on $\left(\mathcal{D}_{x}\cap\mathcal{D}_{y}-\mathcal{D}_{z}\right)$,
and $\left(N_{q}+N_{r}+N_{\breve{x}}+N_{\breve{y}}\right)c$ since
$d\left(\left\{ x\left(k\right)\right\} ,\left\{ y\left(k\right)\right\} \right)$=$c$
on $\left(\mathcal{D}_{x}\cup\mathcal{D}_{y}\right)-\left(\mathcal{D}_{x}\cap\mathcal{D}_{y}\right)$.
Similar decompositions also apply to $q$, and $r$, i.e.
\begin{align*}
p & \triangleq\hat{p}_{N}+\hat{p}_{N_{p}}+\left(N_{q}+N_{r}+N_{\breve{x}}+N_{\breve{y}}\right)c,\\
q & \triangleq\hat{q}_{N}+\hat{q}_{N_{q}}+\left(N_{r}+N_{p}+N_{\breve{y}}+N_{\breve{z}}\right)c,\\
r & \triangleq\hat{r}_{N}+\hat{r}_{N_{r}}+\left(N_{p}+N_{q}+N_{\breve{z}}+N_{\breve{x}}\right)c.
\end{align*}
The following bounds
\begin{align}
\hat{p}_{N} & \leq\hat{q}_{N}+\hat{r}_{N},\label{e:triangle_ineqality_N}\\
p & \leq\hat{p}_{N}+\left(N_{p}+N_{q}+N_{r}+N_{\breve{x}}+N_{\breve{y}}\right)c=\hat{p}_{N}+\left(P-N\right)c\leq Pc,\label{e:upper_bound_p}\\
q & \geq\hat{q}_{N}+\left(N_{r}+N_{p}+N_{\breve{y}}+N_{\breve{z}}\right)c\triangleq\hat{q}_{N}+Q^{\circledcirc}c,\label{e:lower_bound_q}\\
r & \geq\hat{r}_{N}+\left(N_{p}+N_{q}+N_{\breve{z}}+N_{\breve{x}}\right)c\triangleq\hat{r}_{N}+R^{\circledcirc}c,\label{e:lower_bound_r}
\end{align}
and the following identities
\begin{align}
P+Q & =S+N_{\breve{x}}+N_{\breve{y}}+S+N_{\breve{y}}+N_{\breve{z}}=T+N_{\breve{y}}\label{e:P+Q}\\
 & =R+S+2N_{\breve{y}},\label{e:P+Q=00003DR+}\\
Q+R & =S+N_{\breve{y}}+N_{\breve{z}}+S+N_{\breve{x}}+N_{\breve{z}}=T+N_{\breve{z}},\label{e:Q+R}\\
R+P & =S+N_{\breve{x}}+N_{\breve{z}}+S+N_{\breve{x}}+N_{\breve{y}}=T+N_{\breve{x}},\label{e:R+P}\\
N+Q^{\circledcirc}+R^{\circledcirc}-Q & =N+2N_{p}+N_{q}+N_{r}+N_{\breve{x}}+N_{\breve{y}}+2N_{\breve{z}}-N-N_{p}-N_{q}-N_{r}-N_{\breve{y}}-N_{\breve{z}}\nonumber \\
 & =N_{p}+N_{\breve{x}}+N_{\breve{z}},\label{e:Qhole+N-}\\
N+Q^{\circledcirc}+R^{\circledcirc}-P & =N+2N_{p}+N_{q}+N_{r}+N_{\breve{x}}+N_{\breve{y}}+2N_{\breve{z}}-N-N_{p}-N_{q}-N_{r}-N_{\breve{x}}-N_{\breve{y}}\nonumber \\
 & =N_{p}+2N_{\breve{z}},\label{e:Qhole+Rhole}
\end{align}
are required for the proof. Note that so far, all of the variables
we have defined are non-negative.
\par\end{flushleft}

Adopting the above notation, the properties of $d^{\left(c\right)}\left(\cdot,\cdot\right)$
and the triangle inequality for $\tilde{d}\left(\cdot,\cdot\right)$
can be expressed as
\begin{gather}
c\geq p^{\prime},q^{\prime},r^{\prime},\label{e:lower_bound_c}\\
r^{\prime}+q^{\prime}-p^{\prime}\geq0,\label{e:triangle_inequality}\\
\frac{r}{R}+\frac{q}{Q}\geq\frac{p}{P}\Longleftrightarrow PQr+RPq-QRp\geq0.\label{e:triangle_ineqality_previous}
\end{gather}
We need to prove that the triangle inequality holds for the following
three cases (note that the result holds trivially when $\left\{ x\left(k+1\right)\right\} =\left\{ y\left(k+1\right)\right\} =\left\{ z\left(k+1\right)\right\} =\emptyset$):
\begin{enumerate}
\item[(i)] $\left\{ x\left(k+1\right)\right\} =\left\{ y\left(k+1\right)\right\} =\emptyset$,
and $\left\{ z\left(k+1\right)\right\} \neq\emptyset$, i.e. 
\[
\frac{r+c}{R+1}+\frac{q+c}{Q+1}\geq\frac{p}{P}.
\]
\item[(ii)] $\left\{ z\left(k+1\right)\right\} =\emptyset$, and $\left\{ x\left(k+1\right)\right\} \neq\emptyset$
or $\left\{ y\left(k+1\right)\right\} \neq\emptyset$, i.e. 
\[
\frac{r}{R}+\frac{q+c}{Q+1}\geq\frac{p+c}{P+1}\text{ or }\frac{r+c}{R+1}+\frac{q}{Q}\geq\frac{p+c}{P+1}.
\]
\item[(iii)] At least two of $\left\{ x\left(k+1\right)\right\} $, $\left\{ y\left(k+1\right)\right\} $
and $\left\{ z\left(k+1\right)\right\} $ are non-empty, i.e. 
\[
\frac{r+r^{\prime}}{R+1}+\frac{q+q^{\prime}}{Q+1}\geq\frac{p+p^{\prime}}{P+1}.
\]
 
\end{enumerate}
\medskip{}

\noindent \textbf{\textit{Case (i)}}
\begin{align*}
\frac{r+c}{R+1}+\frac{q+c}{Q+1}-\frac{p}{P} & =\frac{P\left(Q+1\right)r+P\left(Q+1\right)c+\left(R+1\right)Pq+\left(R+1\right)Pc-\left(Q+1\right)\left(R+1\right)p}{P\left(Q+1\right)\left(R+1\right)}\\
 & =\frac{P\left(Q+1\right)r+\left(R+1\right)Pc+Pc+\left(Q+R+1\right)Pc-\left(QR+Q+R+1\right)p}{P\left(Q+1\right)\left(R+1\right)}\\
 & =\frac{PQr+RPq-QRp+Pr+Pq+Pc+\left(Q+R+1\right)Pc-\left(Q+R+1\right)p}{P\left(Q+1\right)\left(R+1\right)}\\
 & \geq\frac{P\left(r+q+c\right)+\left(Q+R+1\right)\left(Pc-p\right)}{P\left(Q+1\right)\left(R+1\right)}\\
 & \geq0,
\end{align*}
where we used the triangle inequality \eqref{e:triangle_ineqality_previous}
and the bound $Pc\geq p$ from \eqref{e:upper_bound_p}.

\medskip{}

\noindent \textbf{\textit{Case (ii)}}
\begin{align*}
\frac{r}{R}+\frac{q+c}{Q+1}-\frac{p+c}{P+1} & =\frac{\left(P+1\right)\left(Q+1\right)r+R\left(P+1\right)q+R\left(P+1\right)c-\left(Q+1\right)Rp-\left(Q+1\right)Rc}{\left(P+1\right)\left(Q+1\right)R}\\
 & =\frac{\left(PQ+P+Q+1\right)r+R\left(P+1\right)q-\left(Q+1\right)Rp+R\left(P+1\right)c-\left(Q+1\right)Rc}{\left(P+1\right)\left(Q+1\right)R}\\
 & =\frac{\left(PQ+P+Q+1\right)r+RPq+Rq-QRp-Rp+\left(P-Q\right)Rc}{\left(P+1\right)\left(Q+1\right)R}\\
 & =\frac{PQr+RPq-QRp+\left(P+Q+1\right)r+Rq-Rp+\left(P-Q\right)Rc}{\left(P+1\right)\left(Q+1\right)R}\\
 & \geq\frac{\left(P+Q+1\right)r+Rq-Rp+\left(P-Q\right)Rc}{\left(P+1\right)\left(Q+1\right)R},
\end{align*}
where the last line follows from the triangle inequality \eqref{e:triangle_ineqality_previous}.
Using the bounds \eqref{e:upper_bound_p}, \eqref{e:lower_bound_q},
\eqref{e:lower_bound_r} for $p$, $q$, and $r$, and the identities
$P+Q=R+S+2N_{\breve{y}}$ from \eqref{e:P+Q=00003DR+}, we have
\begin{align*}
 & \left(P+Q+1\right)r+Rq-Rp+\left(P-Q\right)Rc\\
 & \geq\left(R+S+2N_{\breve{y}}+1\right)\left[\hat{r}_{N}+R^{\circledcirc}c\right]+R\left[\hat{q}_{N}+Q^{\circledcirc}c\right]-R\left[\hat{p}_{N}+\left(P-N\right)c\right]+\left(P-Q\right)Rc\\
 & \!\begin{array}[t]{cccc}
= & +\thinspace R\hat{r}_{N} & +\left(S+2N_{\breve{y}}+1\right)\hat{r}_{N} & +R^{\circledcirc}\left(R+S+2N_{\breve{y}}+1\right)c\\
 & +\thinspace R\hat{q}_{N} &  & +Q^{\circledcirc}Rc\\
 & \underbrace{-\thinspace R\hat{p}_{N}} & \underbrace{\text{\qquad\qquad\qquad\qquad}} & \underbrace{-\left(P-N\right)Rc+\left(P-Q\right)Rc}\\
\geq & 0 & +\thinspace0 & +R^{\circledcirc}Rc+Q^{\circledcirc}Rc+\left(N-Q\right)Rc
\end{array}\\
 & =\left[R^{\circledcirc}+Q^{\circledcirc}+N-Q\right]Rc\\
 & =\left[N_{p}+N_{\breve{x}}+N_{\breve{z}}\right]Rc\\
 & \geq0,
\end{align*}
where we used $\hat{r}_{N}+\hat{q}_{N}-\hat{p}_{N}\geq0$ from \eqref{e:triangle_ineqality_N}
and $R^{\circledcirc}+Q^{\circledcirc}+N-Q=N_{p}+N_{\breve{x}}+N_{\breve{z}}$
from \eqref{e:Qhole+N-}.

\medskip{}

\noindent \textbf{\textit{Case (iii)}}
\begin{align}
\frac{r+r^{\prime}}{R+1}+\frac{q+q^{\prime}}{Q+1}-\frac{p+p^{\prime}}{P+1} & =\frac{\left(P+1\right)\left(Q+1\right)\left(r+r^{\prime}\right)+\left(R+1\right)\left(P+1\right)\left(q+q^{\prime}\right)-\left(Q+1\right)\left(R+1\right)\left(p+p^{\prime}\right)}{\left(P+1\right)\left(Q+1\right)\left(R+1\right)}\nonumber \\
 & =\frac{\left(PQ+P+Q+1\right)r+\left(RP+R+P+1\right)q-\left(QR+Q+R+1\right)p}{\left(P+1\right)\left(Q+1\right)\left(R+1\right)}\nonumber \\
 & +\frac{\left(PQ+P+Q+1\right)r^{\prime}+\left(RP+R+P+1\right)q^{\prime}-\left(QR+Q+R+1\right)p^{\prime}}{\left(P+1\right)\left(Q+1\right)\left(R+1\right)}\nonumber \\
 & =\frac{\left(PQr+RPq-QRp\right)+\left(r^{\prime}+q^{\prime}-p^{\prime}\right)}{\left(P+1\right)\left(Q+1\right)\left(R+1\right)}\label{e:proof111-diff1}\\
 & +\frac{\left(P+Q+1\right)r+\left(R+P+1\right)q-\left(Q+R+1\right)p}{\left(P+1\right)\left(Q+1\right)\left(R+1\right)}\label{e:proof111-diff2}\\
 & +\frac{\left(PQ+P+Q\right)r^{\prime}+\left(RP+R+P\right)q^{\prime}-\left(QR+Q+R\right)p^{\prime}}{\left(P+1\right)\left(Q+1\right)\left(R+1\right)}.\label{e:proof111-diff3}
\end{align}
Note that \eqref{e:proof111-diff1} $\geq0$ from the triangle inequalities
\eqref{e:triangle_ineqality_previous} and \eqref{e:triangle_inequality}.
It remains to be shown that \eqref{e:proof111-diff2} + \eqref{e:proof111-diff3}$\geq0$.

Into \eqref{e:proof111-diff2}, we substitute the three identities
$P+Q=T+N_{\breve{y}}$, $Q+R=T+N_{\breve{z}}$ and $R+P=T+N_{\breve{x}}$,
from \eqref{e:P+Q}, \eqref{e:Q+R} and \eqref{e:R+P} respectively.
We also use the upper/lower bounds on $p$, $q$, and $r$, from \eqref{e:upper_bound_p},
\eqref{e:lower_bound_q} and \eqref{e:lower_bound_r} respectively.
This yields the following expression
\begin{align*}
 & \left(P+Q+1\right)r+\left(R+P+1\right)q-\left(Q+R+1\right)p\\
 & \geq\left(T+N_{\breve{y}}+1\right)\left[\hat{r}_{N}+R^{\circledcirc}c\right]+\left(T+N_{\breve{x}}+1\right)\left[\hat{q}_{N}+Q^{\circledcirc}c\right]-\left(T+N_{\breve{z}}+1\right)\left[\hat{p}_{N}+\left(P-N\right)c\right]\\
 & \!\begin{array}[t]{ccccc}
= & +\thinspace\left(T+1\right)\hat{r}_{N} & +\thinspace N_{\breve{y}}\hat{r}_{N} & +\thinspace\left(T+1\right)R^{\circledcirc}c & +\thinspace N_{\breve{y}}R^{\circledcirc}c\\
 & +\thinspace\left(T+1\right)\hat{q}_{N} & +\thinspace N_{\breve{x}}\hat{q}_{N} & +\thinspace\left(T+1\right)Q^{\circledcirc}c & +\thinspace N_{\breve{x}}Q^{\circledcirc}c\\
 & \underbrace{-\thinspace\left(T+1\right)\hat{p}_{N}} & \underbrace{-\thinspace N_{\breve{z}}\hat{p}_{N}} & \underbrace{-\thinspace\left(T+1\right)\left(P-N\right)c} & \underbrace{-\thinspace N_{\breve{z}}\left(P-N\right)c}\\
\geq & 0 & -\thinspace N_{\breve{z}}Nc & +\thinspace\left(T+1\right)\left(R^{\circledcirc}+Q^{\circledcirc}-P+N\right)c & -\thinspace N_{\breve{z}}Pc+N_{\breve{z}}Nc
\end{array}\\
 & =\left[\left(T+1\right)\left(N_{p}+2N_{\breve{z}}\right)-N_{\breve{z}}P\right]c\\
 & \geq\left[\left(T+1\right)N_{\breve{z}}-N_{\breve{z}}P\right]c\\
 & =\left(S+N_{\breve{z}}+1\right)N_{\breve{z}}c,
\end{align*}
where we used $\hat{r}_{N}+\hat{q}_{N}-\hat{p}_{N}\geq0$ from \eqref{e:triangle_ineqality_N},
the bound $-\hat{p}_{N}\geq-Nc$ from \eqref{e:upper_bound_p}, $Q^{\circledcirc}+R^{\circledcirc}-P+N=N_{p}+2N_{\breve{z}}$
from \eqref{e:Qhole+Rhole}, and $T-P=S+N_{\breve{z}}$ from \eqref{e:T}
and \eqref{e:P}.

For \eqref{e:proof111-diff3}, we use $P+Q=T+N_{\breve{y}}$, $Q+R=T+N_{\breve{z}}$
and $R+P=T+N_{\breve{x}}$, from \eqref{e:P+Q}, \eqref{e:Q+R} and
\eqref{e:R+P} respectively, which gives us
\begin{align*}
PQ & =\left(S+N_{\breve{x}}+N_{\breve{y}}\right)\left(S+N_{\breve{y}}+N_{\breve{z}}\right)\\
 & =SS+SN_{\breve{y}}+SN_{\breve{z}}+N_{\breve{x}}S+N_{\breve{x}}N_{\breve{y}}+N_{\breve{x}}N_{\breve{z}}+N_{\breve{y}}S+N_{\breve{y}}N_{\breve{y}}+N_{\breve{y}}N_{\breve{z}}\\
 & =SS+SN_{\breve{x}}+2SN_{\breve{y}}+SN_{\breve{z}}+N_{\breve{x}}N_{\breve{y}}+N_{\breve{y}}N_{\breve{z}}+N_{\breve{z}}N_{\breve{x}}+N_{\breve{y}}N_{\breve{y}},\\
PQ+P+Q & =SS+SN_{\breve{x}}+2SN_{\breve{y}}+SN_{\breve{z}}+N_{\breve{x}}N_{\breve{y}}+N_{\breve{y}}N_{\breve{z}}+N_{\breve{z}}N_{\breve{x}}+N_{\breve{y}}N_{\breve{y}}+T+N_{\breve{y}},
\end{align*}
\begin{align*}
QR & =\left(S+N_{\breve{y}}+N_{\breve{z}}\right)\left(S+N_{\breve{x}}+N_{\breve{z}}\right)\\
 & =SS+SN_{\breve{x}}+SN_{\breve{z}}+N_{\breve{y}}S+N_{\breve{y}}N_{\breve{x}}+N_{\breve{y}}N_{\breve{z}}+N_{\breve{z}}S+N_{\breve{z}}N_{\breve{x}}+N_{\breve{z}}N_{\breve{z}}\\
 & =SS+SN_{\breve{x}}+SN_{\breve{y}}+2SN_{\breve{z}}+N_{\breve{x}}N_{\breve{y}}+N_{\breve{y}}N_{\breve{z}}+N_{\breve{z}}N_{\breve{x}}+N_{\breve{z}}N_{\breve{z}},\\
QR+Q+R & =SS+SN_{\breve{x}}+SN_{\breve{y}}+2SN_{\breve{z}}+N_{\breve{x}}N_{\breve{y}}+N_{\breve{y}}N_{\breve{z}}+N_{\breve{z}}N_{\breve{x}}+N_{\breve{z}}N_{\breve{z}}+T+N_{\breve{z}},
\end{align*}
\begin{align*}
RP & =\left(S+N_{\breve{x}}+N_{\breve{z}}\right)\left(S+N_{\breve{x}}+N_{\breve{y}}\right)\\
 & =SS+SN_{\breve{x}}+SN_{\breve{y}}+N_{\breve{x}}S+N_{\breve{x}}N_{\breve{x}}+N_{\breve{x}}N_{\breve{y}}+N_{\breve{z}}S+N_{\breve{z}}N_{\breve{x}}+N_{\breve{z}}N_{\breve{y}}\\
 & =SS+2SN_{\breve{x}}+SN_{\breve{y}}+SN_{\breve{z}}+N_{\breve{x}}N_{\breve{y}}+N_{\breve{y}}N_{\breve{z}}+N_{\breve{z}}N_{\breve{x}}+N_{\breve{x}}N_{\breve{x}},\\
RP+R+P & =SS+2SN_{\breve{x}}+SN_{\breve{y}}+SN_{\breve{z}}+N_{\breve{x}}N_{\breve{y}}+N_{\breve{y}}N_{\breve{z}}+N_{\breve{z}}N_{\breve{x}}+N_{\breve{x}}N_{\breve{x}}+T+N_{\breve{x}}.
\end{align*}
Hence, expanding \eqref{e:proof111-diff3} results in
\begin{align*}
 & \left(PQ+P+Q\right)r^{\prime}+\left(RP+R+P\right)q^{\prime}-\left(QR+Q+R\right)p^{\prime}\\
 & \footnotesize\begin{array}[t]{ccccccccccc}
= & +\:SSr^{\prime} & +\;\;SN_{\breve{x}}r^{\prime} & +\thinspace2SN_{\breve{y}}r^{\prime} & +\;\;\:SN_{\breve{z}}r^{\prime} & +\thinspace N_{\breve{x}}N_{\breve{y}}r^{\prime} & +\thinspace N_{\breve{y}}N_{\breve{z}}r^{\prime} & +\thinspace N_{\breve{z}}N_{\breve{x}}r^{\prime} & +\thinspace N_{\breve{y}}N_{\breve{y}}r^{\prime} & +\thinspace Tr^{\prime} & +\thinspace N_{\breve{y}}r^{\prime}\\
 & +\thinspace SSq^{\prime} & +\thinspace2SN_{\breve{x}}q^{\prime} & +\;\;\:SN_{\breve{y}}q^{\prime} & +\;\;\:SN_{\breve{z}}q^{\prime} & +\thinspace N_{\breve{x}}N_{\breve{y}}q^{\prime} & +\thinspace N_{\breve{y}}N_{\breve{z}}q^{\prime} & +\thinspace N_{\breve{z}}N_{\breve{x}}q^{\prime} & +\thinspace N_{\breve{x}}N_{\breve{x}}q^{\prime} & +\thinspace Tq^{\prime} & +\thinspace N_{\breve{x}}q^{\prime}\\
 & \underbrace{-\thinspace SSp^{\prime}} & \underbrace{-\;\;SN_{\breve{x}}p^{\prime}} & \underbrace{-\;\;\:SN_{\breve{y}}p^{\prime}} & \underbrace{-\thinspace2SN_{\breve{z}}p^{\prime}} & \underbrace{-\thinspace N_{\breve{x}}N_{\breve{y}}p^{\prime}} & \underbrace{-\thinspace N_{\breve{y}}N_{\breve{z}}p^{\prime}} & \underbrace{-\thinspace N_{\breve{z}}N_{\breve{x}}p^{\prime}} & \underbrace{-\thinspace N_{\breve{z}}N_{\breve{z}}p^{\prime}} & \underbrace{-\thinspace Tp^{\prime}} & \underbrace{-\thinspace N_{\breve{z}}p^{\prime}}\\
\geq & 0 & +\thinspace0 & +\thinspace0 & -\thinspace SN_{\breve{z}}p^{\prime} & +\thinspace0 & +\thinspace0 & +\thinspace0 & -\thinspace N_{\breve{z}}N_{\breve{z}}p^{\prime} & +\thinspace0 & -\thinspace N_{\breve{z}}p^{\prime}
\end{array}\\
 & =-\left(S+N_{\breve{z}}+1\right)N_{\breve{z}}p^{\prime},
\end{align*}
where we used the triangle inequality for $p^{\prime}$, $q^{\prime}$,
$r^{\prime}$ in \eqref{e:triangle_inequality}.

Finally, since $c\geq p^{\prime}$, we can conclude that \eqref{e:proof111-diff2}
+ \eqref{e:proof111-diff3} $\geq0$. $\blacksquare$
\end{document}